**Maxwell viscoelastic sintering kinetics of alumina at hot pressing**

Andrey V. Gusarov*, Pavel A. Podrabinnik, Nestor Washington Solís Pinargote, Ekaterina V. Kuznetsova, Yuri O. Pristinskiy, Tatiana V. Tarasova, Sergey N. Grigoriev

*Moscow State University of Technology STANKIN, Vadkovsky Per. 3a, 127055 Moscow, Russia*

**Abstract.** Recent applications of hot pressing, such as interpenetrating phase composites, require deeper understanding of powder densification kinetics including basic mechanisms and rheology approaches applicable to arbitrary stress states. The present work concerns experimental studying hot-pressing sintering of $\alpha$-alumina submicron powder in the range of temperatures 1300 to 1400 °C and pressures 47 to 80 MPa and aims to reveal mechanisms of densification and develop a robust theoretical framework. Isothermal isobar experiments show that when the porosity becomes below 0.2, it approximately decays as dwell time power -1/2. The densification rate is approximately proportional to the porosity cubed at the porosity below 0.2. The stress exponent of densification is around 3 and independent of temperature. The apparent activation energy of densification is independent of pressure but increases with temperature from approximately 230 kJ/mol at 1300-1350 °C to 900 kJ/mol at 1350-1400 °C. A theoretical approach based on the Maxwell viscoelastic model is developed to describe high-temperature material deformation at arbitrary load. The experimental data are generally consistent with the grain-boundary sliding deformation mechanism. Plastic flow in the domains of stress concentration around pores can explain the increased stress exponent and the reduced apparent activation energy.

* Corresponding author e-mail address: av.goussarov@gmail.com

**Nomenclature**

| | |
|---|---|
| $A$ | arbitrary constant, projected area |
| $B$ | arbitrary constant |
| $C$ | chord length |
| $\mathbf{C}$ | stiffness tensor |
| $D$ | effective grain size |
| $\mathbf{D}$ | compliance tensor |
| $d$ | punch descent |
| $d_0$ | descent due to thermal expansion of the punch |
| $\mathbf{e}$ | strain rate tensor |
| $h$ | height |
| $\mathbf{I}_2$ | second-rank identity tensor |
| $\mathbf{I}_4$ | fourth-rank identity tensor, |
| $K$ | bulk modulus |
| $M$ | number of projection directions |
| $N$ | number of chords |
| $n$ | stress exponent |
| $p$ | pressure |
| $Q$ | activation energy |
| $R$ | universal gas constant |
| $R_c$ | external radius of the plastic domain |
| $R_1$ | cavity radius |
| $R_2$ | external radius |
| $r$ | distance from the center |
| $S$ | cross-section area, surface area |
| $s$ | specific surface of grain boundaries |
| $\mathbf{s}$ | parameter |
| $T$ | temperature |
| $t$ | time |
| $\mathbf{u}$ | flow velocity |
| $V$ | volume |
| $W$ | weight |
| $\mathbf{W}$ | relaxation rate tensor |
| $w$ | reduced densification rate |
| $y$ | arbitrary physical field |

*Greek Symbols*

| | |
|---|---|
| $\alpha$ | linear thermal expansion |
| $\beta$ | extinction coefficient |
| $\delta$ | mean projected area per chord |
| $\eta$ | viscosity |
| $\theta$ | porosity |
| $\mu$ | shear modulus |
| $\nu$ | Poisson's ratio |
| $\rho$ | density |
| $\boldsymbol{\sigma}$ | stress tensor |
| $\sigma_{\parallel}$ | normal stresses in the direction of compression |
| $\sigma_{\perp}$ | normal stresses in the perpendicular direction |
| $\tau_c$ | critical resolved shear stress |
| $\tau_{max}$ | maximum shear stress |
| $\phi$ | sum of porosity and volume fraction of plastic domain |
| $\psi$ | free energy per unit mass |
| $\Omega_n$ | constant |
| $\mathbf{\Omega}$ | unit normal vector |
| $\omega$ | relaxation rate |
| $\omega_c$ | characteristic frequency |

*Subscripts*

| | |
|---|---|
| a | apparent |
| b | bulk |
| d | dry |
| f | final state |
| i | immersed, isostatic |
| *i* | counter |
| *j* | counter |
| o | open pores |
| r | radial direction |
| s | shear, solid |
| *T* | value at temperature *T* |
| u | uniaxial |
| w | wet |
| τ | tangential direction |
| 0 | value at full density |

*Other*

| | |
|---|---|
| $\nabla$ | nabla operator |
| $\cdot$ | dot product, material time derivative |
| $\langle\ \rangle$ | mean |

## 1. Introduction

Hot pressing is a proven technique for compacting powder to obtain high-quality materials of low porosity. Recent advances concern combining hot pressing with additive manufacturing where the compacting sample consists of an additively-manufactured solid structure immersed into powder bed. Zheng *et al.* [1] encapsulated powder sample for hot isostatic pressing into an openwork envelope of the same material 3D-printed by laser powder-bed fusion. Combining different materials of solid lattice and powder for hot pressing results in the so-called interpenetrating phase composites. To synthesize the composites, Zhang *et al.* applied hot-pressing sintering [2] and Alawwa *et al.* used spark plasma sintering [3]. In such works, the focus shifts to chemical and thermomechanical compatibility of components [4]. The flowability of powder becomes important to ensure filling-in multiple internal cavities of solid structures [5].

Stress-strain constitutive relation is quite different in the simultaneously pressed powder and solid domains. The local stress state in the phases is determined not only by the external load conditions of pressing but also by mechanical interaction of the domains. The stress state becomes complex and nonuniform. To understand hot pressing in such conditions, the theoretical rheology approaches to hot-pressing sintering need revising and further development. The present work is restricted by hot pressing of alumina, a promising material for the interpenetrating phase composites [4].

The driving force of conventional sintering is the Laplace pressure arising due to surface tension on curved boundaries of pores. In hot pressing, sintering can be considerably accelerated by application of external pressure much greater than the Laplace one [6]. Such a technique is useful to obtain denser materials with finer grains. The kinetics of hot pressing is generally described by creep models. In the early experiments on hot pressing of high-density alumina with grain size of the order of 100 μm, Coble and Guerard [7] confirmed Nabarro-Herring creep where deformation is controlled by lattice diffusion of aluminum ions with the stress exponent equal to one and the grain-size exponent equal to -2.

Harmer and Brook [8] studied hot pressing of $\alpha$-alumina powders of various grain sizes in the range approximately from 0.3 to 10 μm. The temperatures and pressures of the processes were 1475 to 1630 $^{o}$C and 5 to 20 MPa, respectively. Densification rate was measured at the porosity levels of 0.15 and 0.25. Experiments [8] confirmed diffusion-controlled creep mechanism with the pressure exponent equal to one. The grain-size exponent was -2 at the pressing temperature of 1475 $^{o}$C. At 1630 $^{o}$C, the grain-size exponent varied from -2 at the grain size around one micron to -3 at the grain size around 10 microns. The variation of the grain-size exponent indicates the gradual change of the principal diffusion mechanism from the lattice to boundary one.

Densification of alumina is sensitive to the grain size. Therefore, Roy *et al.* [9] focused on the kinetics of grain growth in the hot pressing process in the ranges of temperature 1450 to 1650 $^{o}$C and pressure 0 to 50 MPa. The initial grain size of the alumina powder was around 0.5 μm. It was found that the grain in the sintered sample can attain the size of few microns even if the process time is as low as 10 minutes and that both the temperature and pressure considerably accelerate grain growth. Preferential growth in the direction perpendicular to the compression axis was also observed [9]. Shen *et al.* [10] published a comprehensive study of alumina grain growth kinetics at isothermal uniaxial pressing in the ranges of temperature 1200 to 1300 $^{o}$C and time 0 to 40 min. They emphasized considerable acceleration of grain growth with temperature and influence of the powder granulometry.

Bernard-Granger and Guizard [11] concluded that hot pressing of alumina is controlled by grain-boundary sliding and oxygen boundary diffusion. They also supposed that grain size can be a state function of porosity [11]. Langer *et al.* [12] studied hot pressing of a submicron alumina powder in the temperature range 1100 to 1200 $^{o}$C and found that densification is controlled by boundary diffusion. They noticed considerable grain growth and pointed out similarity of curves grain size-porosity [12]. Santanach *et al.* [13] identified two sintering regimes at hot pressing of alumina: densification without considerable grain growth at temperatures below 1100-1200 $^{o}$C and grain growth at higher temperatures. Aman *et al.* [14] studied hot pressing of alumina in the temperature range 1150 to 1300 $^{o}$C. Their experiments with constant heating rates contradicted the hypothesis of a universal dependency between the grain size and the porosity proposed in works [11,12].

Antonu *et al.* [15] pointed out similarity between the processes of hot pressing and creep and developed a densification model based on creep deformation. They applied their model to explain isothermal isobar experiments with submicron alumina powder in the temperature range 900 to 1100 $^{o}$C to estimate the stress exponent and the apparent activation energy. It was concluded, that the principal sintering mechanism is grain boundary sliding accompanied with dislocation motion and particle fracture in the considered conditions of initial sintering in the porosity range from 0.25 to 0.4 [15]. Recently, Boldin *et al.* [16] studied sintering kinetics of alumina at 1520 $^{o}$C and 70 MPa. They found that densification is controlled by boundary diffusion at early stages and by grain boundary sliding later [16].

The above review shows that kinetics of alumina densification in hot pressing has been carefully studied at temperatures below 1300 $^{o}$C. Important results have been obtained at temperatures around 1500-1600 $^{o}$C while temperature interval 1300-1400 $^{o}$C is less investigated. However, this interval can be promising for attaining a high density with a moderate grain growth. The present study concerns isothermal isobar experiments at temperatures1300-1400 $^{o}$C and axial compressive stresses 47-80 MPa. The objective is to reveal densification mechanisms at these conditions and develop a robust theoretical framework. Section 2 describes the approach to hot-pressing rheology. Section 3 deals with the applied experimental methods. Section 4 presents experimental results and Section 5 discusses them.

## 2. Rheology of hot pressing

Assume that the Helmholtz free energy of a deformed medium is a non-negative quadratic form of the stress tensor $\boldsymbol{\sigma}$ components,

$$\psi = \frac{\mathbf{D}:\boldsymbol{\sigma}:\boldsymbol{\sigma}}{2\rho} \tag{2.1}$$

where $\psi$ is the free energy per unit mass and the fourth-rank tensor $\mathbf{D}$ depending on density $\rho$ defines the coefficients of the quadratic form. The free energy attains a minimum at $\boldsymbol{\sigma} = 0$. The Clausius-Duhem inequality for isothermal processes is [17]

$$\rho\dot{\psi} \le \boldsymbol{\sigma} : \mathbf{e} \tag{2.2}$$

where strain rate tensor

$$\mathbf{e} = \frac{\nabla\mathbf{u} + (\nabla\mathbf{u})^{\mathrm{T}}}{2}, \tag{2.3}$$

is defined through flow velocity **u** and dot designates material derivative,

$$\dot{y} = \frac{\partial y}{\partial t} + \mathbf{u} \cdot \nabla y, \tag{2.4}$$

where $y$ is an arbitrary physical field, $t$ time, and $\nabla$ nabla operator. Substituting Eq. (2.1) into Eq. (2.2) results

$$\mathbf{D} : \boldsymbol{\sigma} : \dot{\boldsymbol{\sigma}} + \frac{\dot{\rho}}{2}\left(\frac{\mathrm{d}\mathbf{D}}{\mathrm{d}\rho} - \frac{\mathbf{D}}{\rho}\right) : \boldsymbol{\sigma} : \boldsymbol{\sigma} \le \boldsymbol{\sigma} : \mathbf{e}. \tag{2.5}$$

According to the continuity equation,

$$\frac{\dot{\rho}}{\rho} = -\mathbf{I}_2 : \mathbf{e}, \tag{2.6}$$

where $\mathbf{I}_2$ is the second-rank identity tensor. Therefore, Eq. (2.5) becomes

$$\mathbf{D} : \boldsymbol{\sigma} : \dot{\boldsymbol{\sigma}} \le \boldsymbol{\sigma} : \mathbf{e} + \frac{\mathbf{I}_2 : \mathbf{e}}{2}\left(\rho\frac{\mathrm{d}\mathbf{D}}{\mathrm{d}\rho} - \mathbf{D}\right) : \boldsymbol{\sigma} : \boldsymbol{\sigma}. \tag{2.7}$$

The first term of the right-hand side of Eq. (2.7) is linear relative the components of tensors **e** and $\boldsymbol{\sigma}$ while the second term is linear relative the components of **e** but quadratic relative $\boldsymbol{\sigma}$. Therefore, the second term can be neglected at small $\boldsymbol{\sigma}$,

$$\mathbf{D} : \boldsymbol{\sigma} : \dot{\boldsymbol{\sigma}} \le \boldsymbol{\sigma} : \mathbf{e}. \tag{2.8}$$

The relaxation of stresses defines the following Maxwell viscoelastic constitutive relation:

$$\mathbf{D} : (\dot{\boldsymbol{\sigma}} + \mathbf{W} : \boldsymbol{\sigma}) = \mathbf{e}, \tag{2.9}$$

where **W** is the fourth-rank tensor of relaxation rate. If stress relaxation is negligible, deformation becomes reversible and constitutive relation (2.9) reduces to Hooke's law

$$\mathbf{D} : \dot{\boldsymbol{\sigma}} = \mathbf{e}, \tag{2.10}$$

indicating that **D** is the compliance tensor. Equations (2.8) and (2.9) can be double-dot multiplied by stiffness tensor $\mathbf{C} = \mathbf{D}^{-1}$, which is inverse to the compliance tensor, to obtain the equivalent forms of the Clausius-Duhem inequality,

$$\boldsymbol{\sigma} : \dot{\boldsymbol{\sigma}} \le \mathbf{C} : \boldsymbol{\sigma} : \mathbf{e}, \tag{2.11}$$

and the constitutive relation,

$$\dot{\boldsymbol{\sigma}} + \mathbf{W} : \boldsymbol{\sigma} = \mathbf{C} : \mathbf{e}. \tag{2.12}$$

Equation (2.12) can be double-dot multiplied by $\boldsymbol{\sigma}$ and subtracted from inequality (2.11) to reduce the Clausius-Duhem inequality to the following one:

$$\mathbf{W} : \boldsymbol{\sigma} : \boldsymbol{\sigma} \ge 0. \tag{2.13}$$

Thus, the quadratic form corresponding to the relaxation rate tensor **W** should be non-negative. The assumption of smallness of $\boldsymbol{\sigma}$ applied to simplify Eq. (2.7) means that the product of a component of **D** and a component of $\boldsymbol{\sigma}$ is much lower than one. In other words, this assumption is equivalent to the smallness of elastic strain.

The level of stresses can be determined by an external load. If the load is constant or varies with the rate much slower than the relaxation rate, the first term in the left-hand side of Eq. (2.12) can be neglected to obtain the quasi-steady or viscous approximation. The conventional viscous model is obtained by resolving the resulted equation relative the stress tensor,

$$\boldsymbol{\sigma} = \mathbf{H} : \mathbf{e}, \quad \mathbf{H} = \mathbf{W}^{-1}\mathbf{C}, \tag{2.14}$$

where **H** is the viscosity tensor.

In an isotropic medium, the stiffness, relaxation rate, and viscosity tensors are

$$\begin{pmatrix} \mathbf{C} \\ \mathbf{W} \\ \mathbf{H} \end{pmatrix} = \begin{pmatrix} K \\ \omega_\text{b} \\ \eta_\text{b} \end{pmatrix} \mathbf{I}_2\mathbf{I}_2 + 2\begin{pmatrix} \mu \\ \omega_\text{s} \\ \eta_\text{s} \end{pmatrix}\left(\mathbf{I}_4 - \frac{1}{3}\mathbf{I}_2\mathbf{I}_2\right), \tag{2.15}$$

where $\mathbf{I}_4$ is the fourth-rank identity tensor, $K$ the bulk and $\mu$ the shear moduli, $\omega_\text{b}$ the bulk and $\omega_\text{s}$ the shear relaxation rates, and $\eta_\text{b}$ the bulk and $\eta_\text{s}$ the shear viscosities. The first term in the right-hand side of Eq. (2.15) transforms to the trace when double-dot multiplied by a tensor. The second term gives rise to the deviatoric component of a tensor. Coefficients $K$, $\mu$, $\omega_\text{b}$, $\omega_\text{s}$, $\eta_\text{b}$, and $\eta_\text{s}$ should be non-negative to ensure thermodynamic compatibility, in particular non-negativity of quadratic forms (2.1) and (2.13). They are not independent because the right Eq. (2.14) defines the following constraints:

$$\eta_\text{b} = \frac{K}{3\omega_\text{b}}, \quad \eta_\text{s} = \frac{\mu}{2\omega_\text{s}}. \tag{2.16}$$

Bulk relaxation is allowed in porous media. When pores vanish and full density $\rho_0$ is attained, isotropic deformation becomes reversible. Therefore, bulk relaxation rate $\omega_\text{b}$ tends to zero if density $\rho$ tends to $\rho_0$.

The conventional processes of hot pressing are isostatic pressing with pressure $p$ and uniaxial pressing with zero strain in the directions perpendicular to the compression axis. At isostatic pressing, the deviatoric components of the strain rate and stress tensors vanish. Any normal stress component is replaced with $-p$ in the left-hand side of Eq. (2.12). In the right-hand side, the trace of the strain rate tensor is related to the density by the continuity equation (2.6). Thus, constitutive relation (2.12) reduces to the following pressure-density relation:

$$\dot{p} + 3\omega_\text{b} p = K\frac{\dot{\rho}}{\rho}. \tag{2.17}$$

At uniaxial pressing, the stress state is specified by normal stresses $\sigma_{||}$ in the direction of compression and $\sigma_\perp$ in the perpendicular direction. One can obtain the same Eq. (2.17) for the hydrostatic pressure

$$p = -\frac{\sigma_{||} + 2\sigma_\perp}{3}, \tag{2.18}$$

and the following equation for the difference of the normal stress components:

$$(\dot{\sigma}_\perp - \dot{\sigma}_{||}) + 2\omega_\text{s}(\sigma_\perp - \sigma_{||}) = 2\mu\frac{\dot{\rho}}{\rho}. \tag{2.19}$$

Differential equations (2.18) and (2.19) describe kinetics of sintering. Functional dependencies between their coefficients are considered below.

The elastic constants of low-porous media are commonly estimated by Mackenzie relations [18]:

$$\frac{1}{K}=\frac{\rho_0}{K_0\rho}+\frac{3}{4\mu_0}\left(\frac{\rho_0}{\rho}-1\right),\qquad \frac{\mu_0-\mu}{\mu_0}=5\frac{3K_0+4\mu_0}{9K_0+8\mu_0}\left(1-\frac{\rho}{\rho_0}\right),\tag{2.20}$$

where index 0 designates values in the non-porous medium with full density. The left Eq. (2.20) can also be obtained from the model problem of elastic isostatic compression of a spherical layer [12,13] with an arbitrary ratio of external and internal radii. Therefore, this equation is reasonable for high-porous media too. On the contrary, the right Eq. (2.20) can take negative values with decreasing $\rho$ and is not acceptable at high porosity. For high-porous media, Skorohod [20] proposed a differential approach. According to this approach, difference $\rho-\rho_0$ is replaced with differential d$\rho$ and $\rho_0$ with $\rho$. The same for all variables. Thus, the right Eq. (2.20) results in differential equation

$$\frac{\mathrm{d}\mu}{\mu}=\left(5\frac{3K+4\mu}{9K+8\mu}\right)\frac{\mathrm{d}\rho}{\rho}.\tag{2.21}$$

An approximate analytical solution of Eq. (2.21) is considered elsewhere [20].

The viscosity law (2.14) is similar to Hooke's law. Therefore, the Mackenzie relations are applicable to the bulk and shear viscosities given by Eq. (2.16). Namely, bulk modulus $K$ is replaced with bulk viscosity $\eta_b$ and shear modulus $\mu$ with shear viscosity $\eta_s$ in Eq. (2.20). One should account for vanishing bulk relaxation at full density. Thus, the following relations for the relaxation rates in low-porous media are obtained:

$$\frac{\omega_b}{K}=\frac{\omega_{s0}}{2\mu_0}\left(\frac{\rho_0}{\rho}-1\right),\qquad \frac{\omega_s}{\mu}=\frac{\omega_{s0}}{\mu_0}\frac{3\rho_0}{5\rho-2\rho_0}.\tag{2.22}$$

Appendix A considers the model problem of viscoelastic isotropic compression of a spherical layer. The same equation for the bulk relaxation rate is obtained there at an arbitrary value of density. Therefore, the left Eq. (2.22) can be applied not only to low-porous but also high-porous media. The differential approach for the right Eq. (2.22) results in the following differential equation:

$$\frac{\mu}{\omega_s}\mathrm{d}\left(\frac{\omega_s}{\mu}\right)=-\frac{5}{3}\frac{\mathrm{d}\rho}{\rho},\tag{2.23}$$

with the exact analytical solution for the shear relaxation rate applicable at any value of density

$$\frac{\omega_s}{\mu}=\frac{\omega_{s0}}{\mu_0}\left(\frac{\rho_0}{\rho}\right)^{\frac{5}{3}}.\tag{2.24}$$

In isostatic pressing at constant pressure $p$, the first term in the left-hand side of Eq. (2.17) vanishes. Substitution of Eq. (2.22) for the bulk relaxation rate results in the following kinetic equation:

$$\frac{\dot\theta}{\theta}=-\frac{3\omega_{s0}p}{2\mu_0},\qquad \theta=\frac{\rho_0-\rho}{\rho_0},\tag{2.25}$$

written in terms of porosity $\theta$. In uniaxial pressing at constant axial stress $\sigma_{||}$, Eq. (2.17) should be solved together with Eqs. (2.18) and (2.19). Hydrostatic pressure (2.18) is no longer constant. However, material derivatives in the left-hand sides of Eqs. (2.17) and (2.19) can still be neglected if the stresses are much lower than the elastic moduli. This is the quasi-steady approximation. In this approximation, the system of Eqs. (2.17)-(2.19) reduces to the following kinetic equation:

$$\left[\frac{1}{\theta}+(1-\theta)^{\frac{2}{3}}\right]\dot{\theta}=\frac{3\omega_{s0}\sigma_{\|}}{2\mu_0}. \tag{2.26}$$

Hydrostatic pressure $p$ depends on porosity as

$$-\frac{\sigma_{\|}}{p}=1+\theta(1-\theta)^{\frac{2}{3}}. \tag{2.27}$$

Equation (2.27) indicates that hydrostatic pressure $p$ is not greater than axial pressure $-\sigma_{\|}$ in axial pressing. The additional term of $(1-\theta)^{2/3}$ in the left-hand side of Eq. (2.26) describes the decrease of the densification rate due to the decrease of the hydrostatic pressure.

The obtained kinetic equations (2.25) and (2.26) are similar. The difference is due to the different stress states at isostatic and uniaxial pressing. The right-hand side of these equations is proportional to shear relaxation rate $\omega_{s0}$, which is generally a function of temperature. One can introduce a dependence of the relaxation rate on stress components to obtain non-linear models of creep commonly applied to describe hot pressing. It is known that creep is sensitive to the microstructure. Therefore, $\omega_{s0}$ should depend on parameters characterizing the material microstructure, for example the mean grain size. The microstructure evolution is generally coupled with the densification kinetics.

## 3. Experimental methods

### *3.1. Preparation and characterization of initial powder*

Commercially available $Al_2O_3$ powder (Taimei Chem. Co. Ltd., Tokyo, Japan) is used. The raw powder contains non-equilibrium γ-phase. Phase transformation γ-α could influence densification kinetics during hot pressing. To exclude this influence, the powder is annealed at 1400 °C for two hours. Thus, the γ-α transformation is accomplished before the hot pressing starts. Figure 1 shows the annealed powder. It consists of particles essentially smaller than 100 microns in size, see Fig. 1a. Figure 1b shows the surface of a particle. One can see that the particle consists of submicron grains with pores between them. An X-ray diffraction study confirms that the annealed powder is essentially α $Al_2O_3$ because the spectrum shown in Fig. 1c contains the principal peaks of α phase and no peaks of γ phase.

### *3.2. Hot pressing*

The powder is poured into the space limited with a graphite matrix with a cylindrical hole and two graphite cylindrical punches as shown in Fig. 2. The sample is cold-pressed with the axial pressure of 40 MPa and then placed into hot press SPS H-HP-D25-SD (FCT Systeme GmbH, Rauenstein, Germany) where it is heated at the rate of 100 K/min up to the dwell temperature, isothermally pressed with a constant axial load, and, finally, cooled down to the room temperature with the press. Eighteen samples of $Al_2O_3$ ceramics are obtained at pressures 47 and 80 MPa, temperatures 1300, 1350, and 1400 °C, and dwell times up to 120 min as indicated in Table 1.

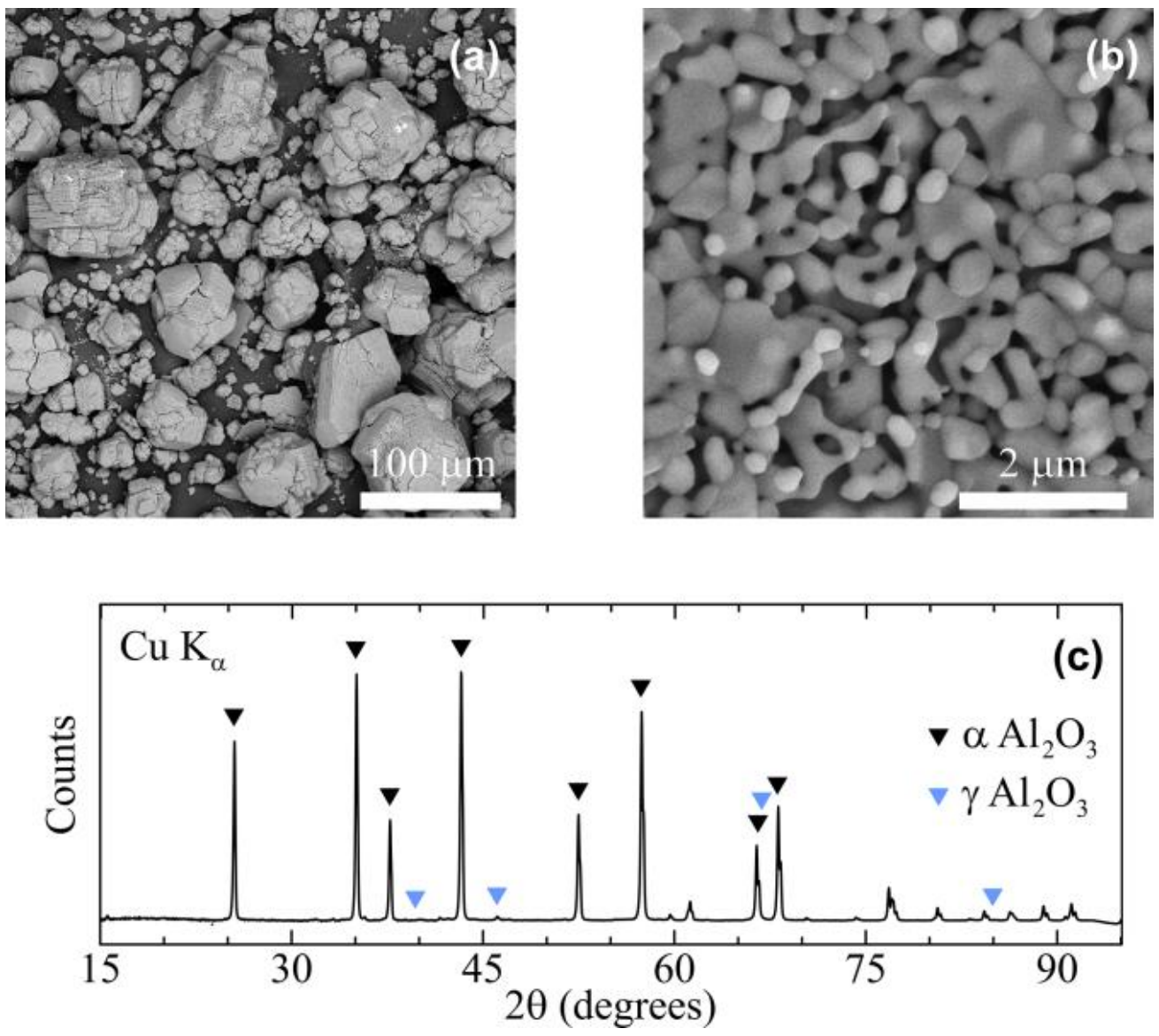

**Fig. 1.** Initial powder: general view (a), surface of a particle at a higher magnification (b), and X-ray diffraction spectrum (c).

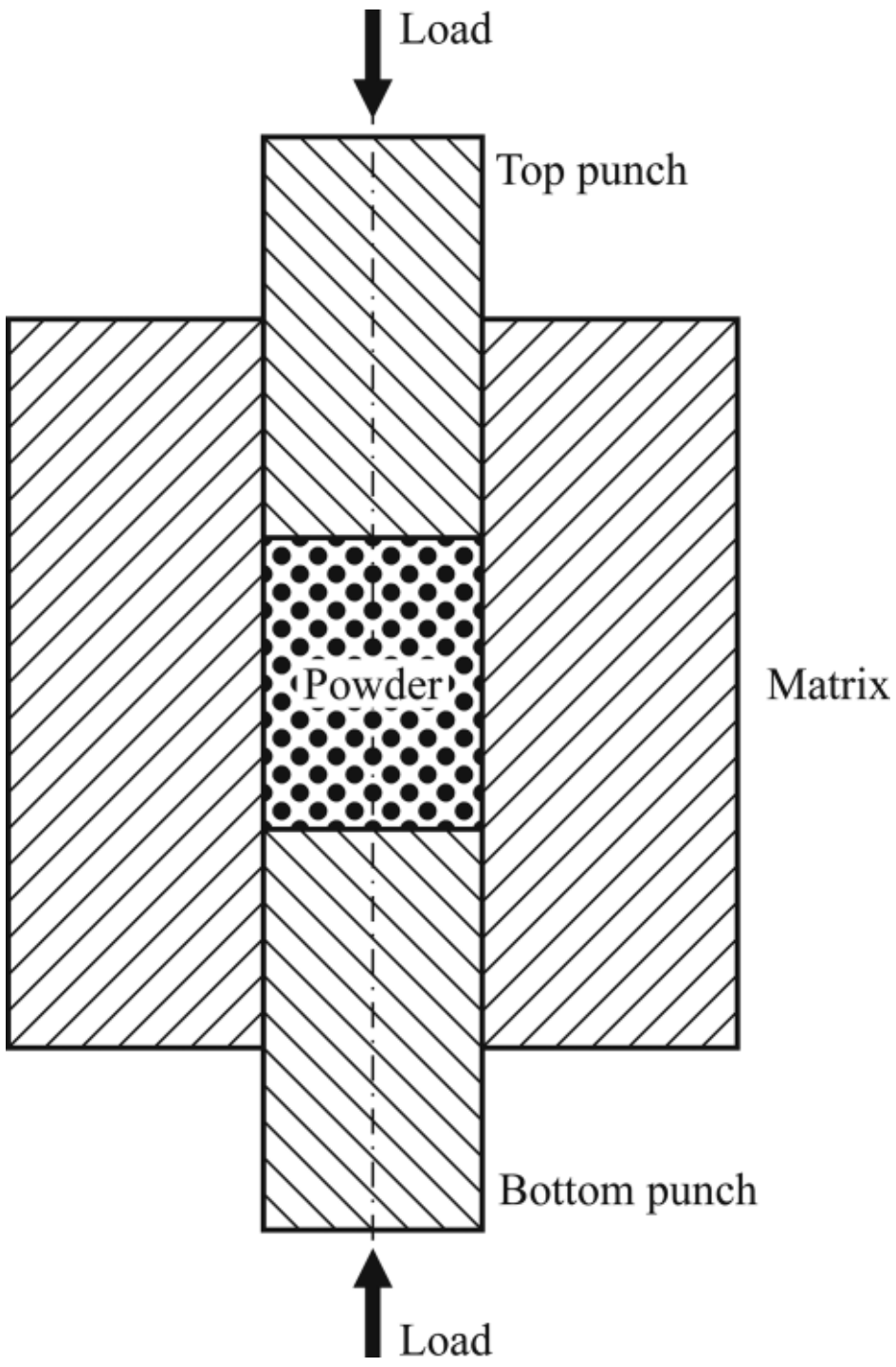

**Fig. 2.** Hot pressing.

**Table 1.** Parameters of hot pressing.

| | | Axial pressure | |
|---|---|---|---|
| | | 47 MPa | 80 MPa |
| Temperature | 1300 °C | Dwell: 15, 30, 60, and 120 min | Dwell: 15, 30, 60, and120 min |
| | 1350 °C | No samples | Dwell: 15, 30, 60, and 120 min |
| | 1400 °C | Dwell: 15, 30, and 60 min | Dwell: 15, 30, and 60 min |

*3.3. Process characterization*

The principal characteristics, such as temperature, pressing force, and punch descent are recorded during the hot-pressing process. Temperature $T$ is measured by a pyrometer focused on the bottom of a hole drilled in the top punch. The absolute value of axial normal stress $\sigma_{||}$ is calculated as the ratio of the pressing force to the punch cross section area. Punch descent $d$ is the decrease of the sample height during pressing. The final value of $d$ at the end of the isothermal dwell is designated $d_{\mathrm{f}}$. After cooling down to the room temperature, the sample is taken from the press. Sample density is measured by the Archimedes's method to estimate final porosity $\theta_{\mathrm{f}}$. Final sample height $h_{\mathrm{f}}$ is measured by a micrometer.

The porosity of the hot-compressing sample is defined as

$$\theta = \frac{V_T - V_{\mathrm{s}T}}{V_T}, \tag{3.1}$$

where $V_T$ is the sample volume and $V_{\mathrm{s}T}$ the volume occupied by the solid phase at pressing temperature $T$. Assume that the process is isothermal, the hydrostatic pressure is much less than the bulk modulus, and the mass loss is negligible. Then, $V_{\mathrm{s}T}$ is a constant, which can be obtained from Eq. (3.1) applied to the final compression state characterized by the final porosity $\theta_{\mathrm{f}}$ and height $h_{\mathrm{f}T}$ and cross-section area $S_T$ of the sample at the pressing temperature,

$$\theta_{\mathrm{f}} = \frac{h_{\mathrm{f}T} S_T - V_{\mathrm{s}T}}{h_{\mathrm{f}T} S_T}. \tag{3.2}$$

It is assumed that thermal expansion of the solid and the void phases is congruent. Therefore, the value of porosity $\theta_{\mathrm{f}}$ is the same at the pressing temperature and after cooling down to the room temperature. On the contrary, the height of the sample measured at the room temperature is

$$h_{\mathrm{f}} = \frac{h_{\mathrm{f}T}}{1+\alpha}, \tag{3.3}$$

where $\alpha$ is the linear thermal expansion in the interval from the room temperature to the pressing temperature. The cross section of the sample is restricted by the cross section of the matrix. Therefore, $S_T$ is constant in the considered isothermal processes.

Let function $d(t)$ be the descent of the punch versus time $t$ and $d_{\mathrm{f}}$ the final value of this function at the end of isothermal compression. Then, the height of the sample is

$$h_T = h_{\mathrm{f}T} + d_{\mathrm{f}} - d, \tag{3.4}$$

in isothermal compression. The temporal dependence of porosity is obtained from Eq. (3.1) as

$$\theta = \frac{h_T S_T - V_{\mathrm{s}T}}{h_T S_T}. \tag{3.5}$$

Three unknowns, $h_T$, $h_{\mathrm{f}T}$, and ratio $V_{\mathrm{s}T}/S_T$, can be excluded from four equations (3.2)-(3.5) to obtain the following expression for the porosity:

$$\theta = 1 - \frac{(1-\theta_{\mathrm{f}})(1+\alpha)h_{\mathrm{f}}}{(1+\alpha)h_{\mathrm{f}} + d_{\mathrm{f}} - d}. \tag{3.6}$$

The temporal derivative of Eq. (3.6) is

$$\dot{\theta} = -\frac{(1-\theta_\mathrm{f})(1+\alpha)h_\mathrm{f}}{\left[(1+\alpha)h_\mathrm{f} + d_\mathrm{f} - d\right]^2}\dot{d} \,. \tag{3.7}$$

The value of $\dot{d}$ in the right-hand side of Eq. (3.7) is calculated by numerical differentiation of data set $d(t)$ captured by the press.

To exclude possible thermal expansion of the punch, the process is repeated without powder. Then, the obtained function $d_0(t)$ corresponds to the thermal expansion of the punch and the corrected descent of the punch is

$$d \to d - d_0 \,. \tag{3.8}$$

The present work has shown that correction (3.8) is significant during heating and cooling but essentially negligible at the isothermal dwell in the studied conditions.

Final porosity $\theta_\mathrm{f}$ is measured by the Archimedes' method. The sintered sample is dried and weighted in air to obtain its dry weight $W_\mathrm{d}$. Then, the sample is immersed into water and the immersed weight $W_\mathrm{i}$ is recorded when it attains a steady value. Finally, the sample is weighted in air to obtain wet weight $W_\mathrm{w}$. The volume of open pores filled with water is

$$V_\mathrm{o} = \frac{W_\mathrm{w} - W_\mathrm{d}}{\rho_\mathrm{H2O}} \,, \tag{3.9}$$

where $\rho_\mathrm{H2O}$ is the density of water. Apparent volume of the sample $V_\mathrm{a}$ should be corrected for the volume of open pores,

$$V_\mathrm{a} - V_\mathrm{o} = \frac{W_\mathrm{d} - W_\mathrm{i}}{\rho_\mathrm{H2O}} \,. \tag{3.10}$$

The apparent density of the sample is

$$\rho = \frac{W_\mathrm{d}}{V_\mathrm{a}} = \frac{W_\mathrm{d}}{W_\mathrm{w} - W_\mathrm{i}}\rho_\mathrm{H2O} \,. \tag{3.11}$$

The total porosity including the open and closed ones is

$$\theta_\mathrm{f} = 1 - \frac{\rho}{\rho_\mathrm{Al2O3}} \,, \tag{3.12}$$

where $\rho_\mathrm{Al2O3}$ is the full density of $Al_2O_3$.

The values of water and $Al_2O_3$ densities at the room temperature and linear thermal expansion of $Al_2O_3$ are taken from literature [21-23] and listed in Table 2.

**Table 2.** Accepted values of density and thermal expansion

| Parameter | Temperature | Value | Reference |
|---|---|---|---|
| Density of $H_2O$ | 25 °C | $\rho_\mathrm{H2O}$ = 0.997 g/cm$^3$ | [21] |
| Density of $Al_2O_3$ | 25 °C | $\rho_\mathrm{Al2O3}$ = 3.987 g/cm$^3$ | [22] |
| Linear thermal expansion of $Al_2O_3$ from 25 °C | 1300 °C | $\alpha$ = 1.11% | [23] |
| | 1350 °C | $\alpha$ = 1.16% | |
| | 1400 °C | $\alpha$ = 1.21% | |

*3.4. Microstructure characterization*

The samples are cut, ground, and polished, then thermally etched at the temperature 150 K below the pressing temperature for two hours to reveal microstructure. The cross-section images are obtained with VEGA 3 LMH scanning electron microscope (Tescan, Brno, Czech Republic) in the secondary-electron mode. The images are analyzed by the chord method to calculate the specific surface and effective grain size. The quality of images is insufficient for automatic processing. Therefore, grain boundaries are manually drawn over the images. Chord-length distributions are obtained using Point Tools plugin of ImageJ software.

Chord analysis is based on Cauchy's theorem about the quadrature of curved surfaces (Theorem IV) [24] stating that the half of the surface area $S$ is equal to the mean projected area $A$ of the surface on a plane specified by an arbitrary unit normal vector $\mathbf{\Omega}$,

$$\frac{S}{2} = \langle A \rangle = \frac{1}{4\pi} \int_{4\pi} A(\mathbf{\Omega}) \mathrm{d}\mathbf{\Omega} , \qquad (3.13)$$

where $\mathrm{d}\mathbf{\Omega}$ is the element of solid angle. The projected area is averaged over all possible orientations of plane normal $\mathbf{\Omega}$. Figure 3a shows an example for calculating the projected area of a grain in direction $\mathbf{\Omega}$. In the example, one can split the surface of a grain into four branches $S_1$ .. $S_4$ giving rise four projections $A_1$ .. $A_4$. The total projected area of the grain is

$$A = A_1 + A_2 + A_3 + A_4 . \qquad (3.14)$$

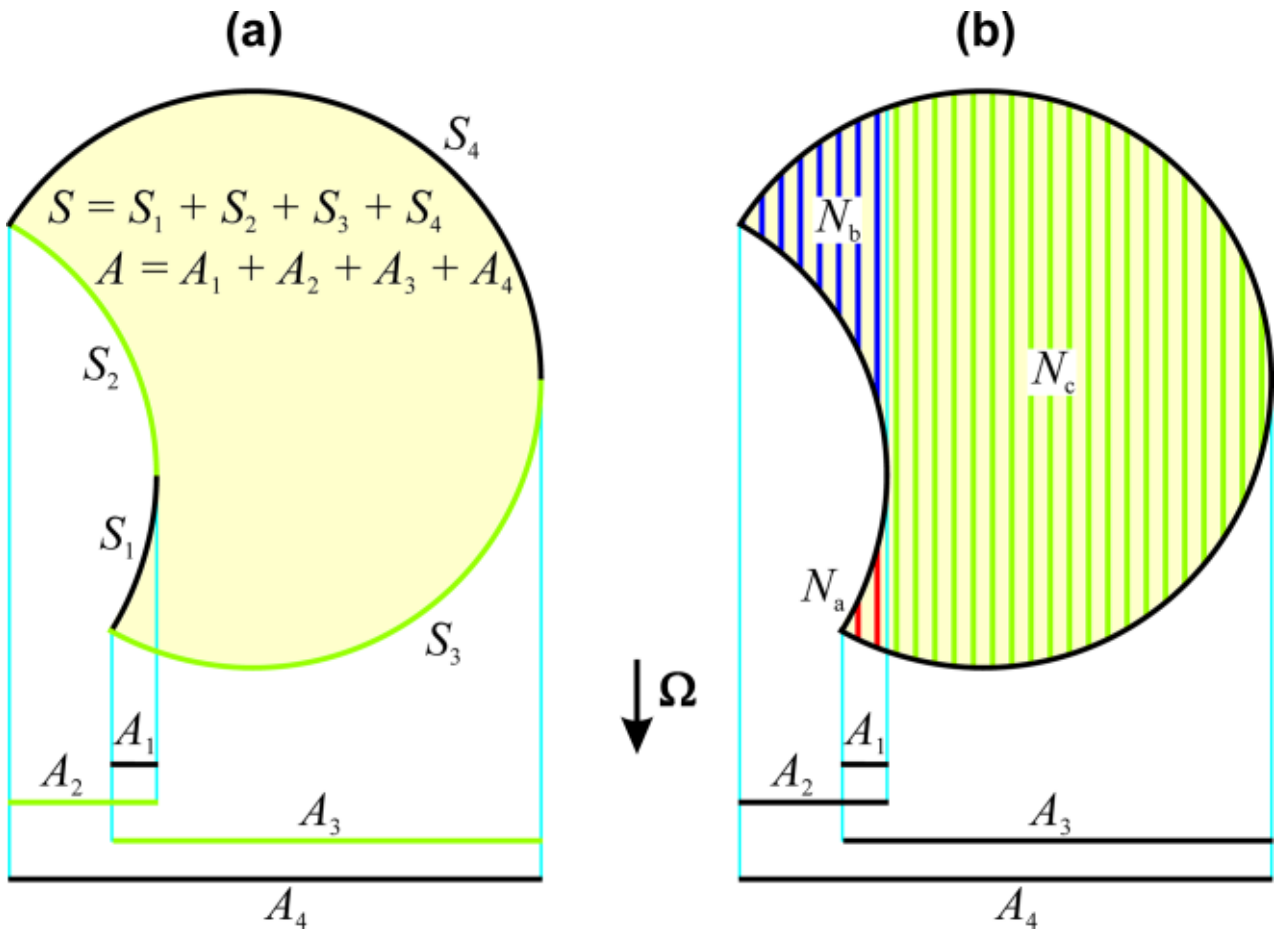

**Fig. 3.** Projected area of a grain in direction $\mathbf{\Omega}$: splitting the grain surface and projection into branches (a) and estimating the projected area by chords parallel to vector $\mathbf{\Omega}$.

Cover the body of the grain with a great number of chords in direction $\mathbf{\Omega}$ as shown in Fig. 3b. Let the parallel chords are uniformly distributed in space with the mean projected area $\delta$ per chord. There are three groups of chords in Fig. 3b drawn by different colors. The number of chords in groups "a", "b", and "c" are $N_a$, $N_b$, and $N_c$, respectively. The total number of chords is

$$N = N_a + N_b + N_c . \qquad (3.15)$$

One can see from Fig 3a that chords of group "a" project on area $A_1$. This area is estimated as

$$A_1 = N_a \delta . \quad (3.16)$$

Similarly,

$$A_2 = N_b \delta \text{, and} \quad (3.17)$$

$$A_3 - A_1 = A_4 - A_2 = N_c \delta . \quad (3.18)$$

Equations (3.14)-(3.18) result in the following estimate for projected area:

$$A = 2N\delta , \quad (3.19)$$

where the values of *A* and *N* depend on direction **Ω**.

In the chord analysis, a great number *M* of projection directions $\mathbf{\Omega}_i$ is used with index *i* ranging from 1 to *M*. The distribution of the projection directions is isotropic. Therefore, integration in Eq. (3.13) can be approximately replaced by averaging as

$$\frac{S}{2} = \frac{1}{M}\sum_{i=1}^{M} A_i = \frac{2\delta}{M}\sum_{i=1}^{M} N_i , \quad (3.20)$$

where $A_i$ is the projection area and $N_i$ the number of chords covering the grain in direction $\mathbf{\Omega}_i$. The volume of the grain is estimated by the rectangle method as

$$V = \delta \sum_{j=1}^{N_i} C_{ij} = \frac{\delta}{M}\sum_{i=1}^{M}\sum_{j=1}^{N_i} C_{ij} , \quad (3.21)$$

where $C_{ij}$ is the length of *j*-th chord in direction $\mathbf{\Omega}_i$. Equations (3.20) and (3.21) result in the following estimate for specific surface *S*/*V* of the grain:

$$\frac{S}{V} = 4\frac{\sum_{i=1}^{M} N_i}{\sum_{i=1}^{M}\sum_{j=1}^{N_i} C_{ij}} . \quad (3.22)$$

The numerator of the right-hand side is the total number of chords while the denominator is their total length. Thus, the specific surface is reciprocal to the mean chord length <C>.

In a monophase material, it is not useful to account for the grain surface area but the area of grain boundaries. Every element of grain boundary separates two grains. Therefore, the area of grain boundaries is a half of the grain surface area and the expression for the specific surface of grain boundaries becomes

$$s = \frac{2}{\langle C \rangle} . \quad (3.23)$$

It is clear that mean grain size is about mean chord <C>. Suppose that the grains are spheres of diameter *D*, surface area $S = \pi D^2$, and volume $V = \pi D^3/6$. Then, Eq. (3.22) results

$$D = \frac{3}{2}\langle C \rangle . \quad (3.24)$$

Below, this value is accepted as an effective grain size.

## 4. Experimental results

Six isothermal isobar processes are studied. Table 3 shows process parameters along with the characteristics of the sintered samples. Figure 4 shows the recorded functions of temperature $T$ and punch descent $d$ versus time $t$. The measured dwell temperature (see plateau of red curves in Fig. 4) and dwell time (plateau length) are equal to the corresponding nominal values indicated in Table 3. The measured heating rate (see the red curves before the temperature plateau) is equal to the specified 100 K/min while the measured cooling rate (see the red curves after the temperature plateau) is greater than 100 K/min. The maximum value attained by the black curve is recorded in Table 3 as final punch descent $d_f$.

**Table 3.** Hot pressing processes

| Process | a | d | b | e | c | f |
|---|---|---|---|---|---|---|
| Axial pressure, $-\sigma_{\|}$ (MPa) | 47 | 80 | 47 | 80 | 47 | 80 |
| Dwell temperature (°C) | 1300 | 1300 | 1350 | 1350 | 1400 | 1400 |
| Dwell time (min) | 120 | 120 | 120 | 120 | 60 | 60 |
| Sample porosity, $\theta_f$ (mm) | 0.16416 | 0.07383 | 0.12852 | 0.05829 | 0.06636 | 0.03052 |
| Sample height, $h_f$ (mm) | 15.33±0.08 | 13.67±0.15 | 13.98±0.02 | 12.61±0.10 | 13.10±0.06 | 12.28±0.09 |
| Final punch descent, $d_f$ (mm) | 12.4 | 15.8 | 12.7 | 15.8 | 14.7 | 17.2 |

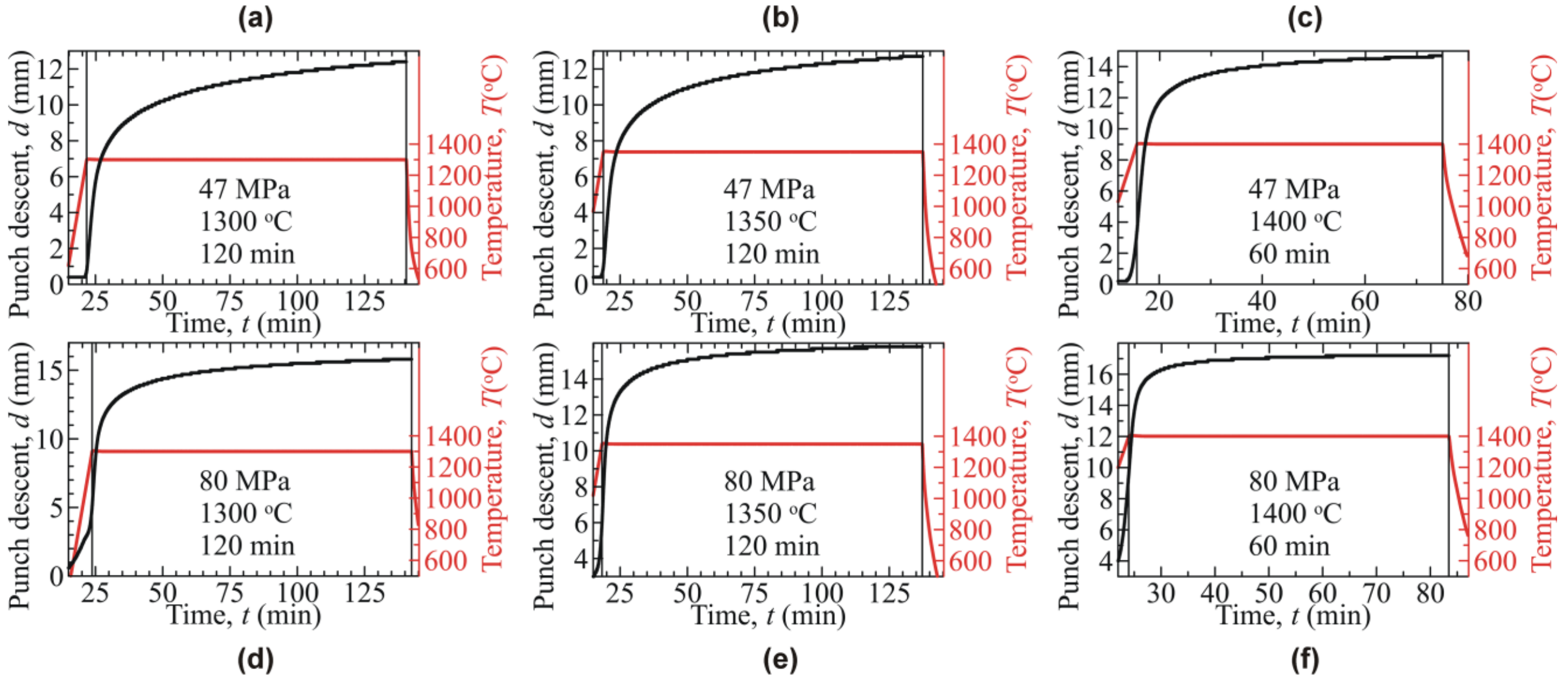

**Fig. 4.** Recorded sample temperature $T$ and punch descent $d$ versus time $t$. Nominal process parameters are indicated in the middle of each diagram: axial pressure $-\sigma_{\|}$, dwell temperature, and dwell time.

### 4.1. *Sintering kinetics*

The recorded data for dependence $d(t)$ shown in Fig. 4 are the input to restore porosity $\theta(t)$ by Eq. (3.6). The values of $\theta_f$, $h_f$, and $d_f$ are taken from Table 3. The values of $\alpha$ are taken from Table 2. Figure 5 shows that the porosity decreases with time $t$. It is reasonable that the rate of decrease raises with both the pressure and the temperature. When the porosity becomes below 0.2, it approximately follows power law with exponent -1/2 shown by dashed lines in Fig. 5. Numerical differentiation of initial data $d(t)$ results in porosity time derivative $\dot{\theta}(t)$ according to Eq. (3.7) and densification rate

$$\frac{\dot{\rho}}{\rho} = \frac{\dot{\theta}}{1-\theta}. \tag{4.1}$$

Figure 6 reveals a strong correlation between the densification rate and porosity $\theta$. In an isothermal isobar process, a power-law dependence versus $\theta$ is clear with the exponent around 3 below the porosity values around $\theta$ = 0.2 .. 0.3. Dashed lines show cubic fits

$$\frac{\dot{\rho}}{\rho} = w\theta^3, \tag{4.2}$$

with factor $w$ strongly dependent on the temperature and pressure. Table 4 shows the factors derived from Fig. 6. The coefficients characterize densification rate at the similar structural states with the same porosity. They increase both with temperature and pressure.

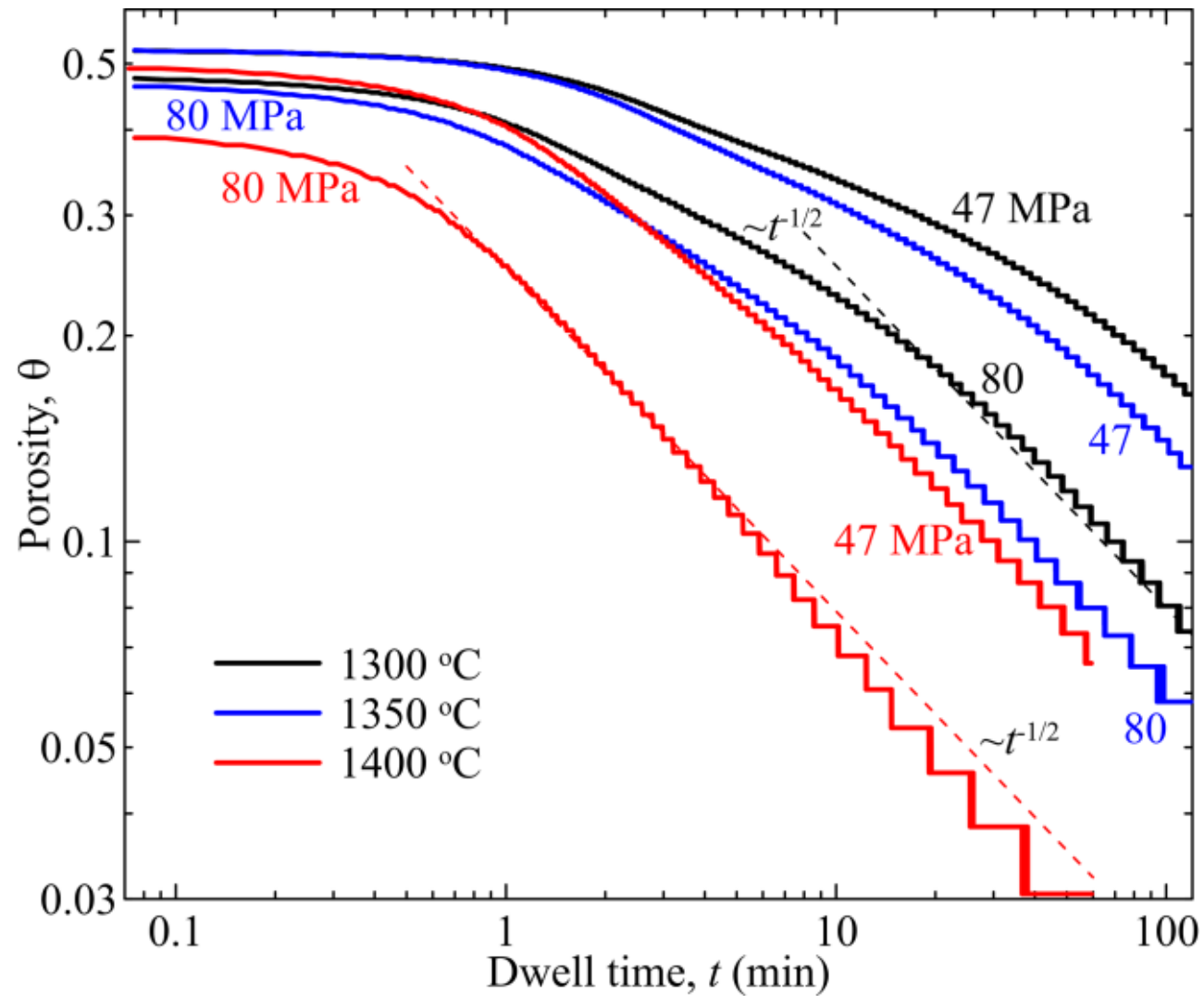

**Fig. 5.** Restored evolution of porosity $\theta$. The dashed lines show power-law decrease.

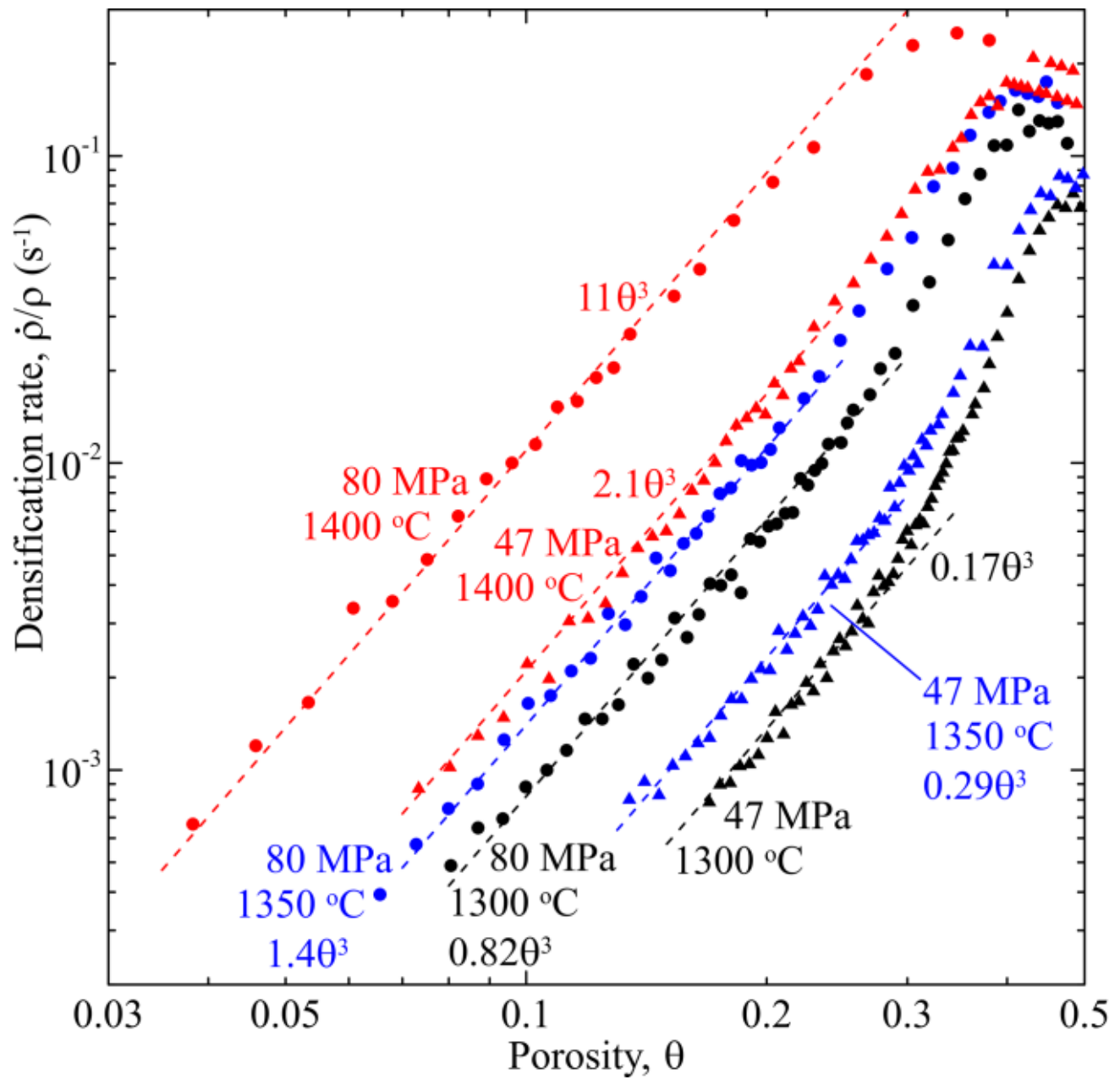

**Fig. 6.** Correlation between densification rate and porosity. The dashed lines show quadratic-law fits.

**Table 4.** Factors *w* ($s^{-1}$) of cubic fit by Eq. (4.2) for isothermal isobar processes

| | | Temperature | | |
|---|---|---|---|---|
| | | $T$ = 1300 °C | $T$ = 1350 °C | $T$ = 1400 °C |
| Axial pressure | $-\sigma_{\|}$ = 47 MPa | 0.17 | 0.29 | 2.1 |
| | $-\sigma_{\|}$ = 80 MPa | 0.82 | 1.4 | 11 |

*4.2. Microstructure*

Figure 7 shows representative SEM micrographs of thermally etched polished cross sections of alumina samples prepared at different temperatures, axial pressures, and isothermal dwell times. The following thermal etching revealed grain boundaries. All the samples exhibit a polycrystalline microstructure consisting of polyhedral grains. Residual pores are found mainly along grain boundaries and at triple junctions. Most samples contain both equiaxed and elongated grains. Elongated grains were most clearly discernible in specimens sintered at 1300 °C and 47 MPa (Figs. 7a-d). The microstructure of samples sintered at 1300 °C and 80 MPa (Figs. 7e-h) is predominantly equiaxed. In addition, the images reveal numerous regions in which the long axes of elongated grain sections are inclined at similar angles to the pressing axis (Fig. 8). Comparable morphological anisotropy in $Al_2O_3$ ceramics processed under uniaxial pressure was previously demonstrated by comparing sections parallel and perpendicular to the compression axis [25]. Hanyu *et al.* [26] noted that under homogeneous uniaxial compression, the maximum shear stress acts on planes inclined at 45° to the loading axis. Given the potential spatial variations in the local stress field [27], the planes of maximum shear may lie at angles other than 45° to the pressing axis (Fig. 8).

The samples sintered at 1300 and 1350 °C (Fig. 7) exhibit a homogeneous microstructure with grain boundaries appeared as thin lines. The highest residual porosity is observed at 1300 °C and 47 MPa, where fine pores occur along grain boundaries and form clusters in some regions. Increasing the pressure to 80 MPa and the temperature to 1350 °C produces a denser microstructure with mostly isolated pores and occasional local clusters. This observation is consistent with the densification curves shown in Fig. 5. Although thermal etching reveals grain boundaries, the clarity and continuity of the boundary traces in the micrographs varied considerably among the samples. During thermal etching, surface diffusion redistributes material along the polished surface, producing local relief where grain boundaries intersect the surface [27,28]. Variations in this relief may partly account for the observed differences in grain-boundary contrast.

To prepare the images for quantitative analysis, grain boundaries are manually traced as vector paths on the original micrographs (Fig. 9a). A closed contour is drawn when the boundary could be followed around the entire grain. The resulting grain-boundary network is overlaid on the corresponding micrograph to visually verify that the contours follow the observed boundary traces (Fig. 9b). Pores and grains intersecting the image edges are filled in black to exclude them from the measurements. The resulting grain map is exported as a separate image aligned with the corresponding micrograph and imported into the ImageJ sowtware for quantitative analysis (Fig. 9c).

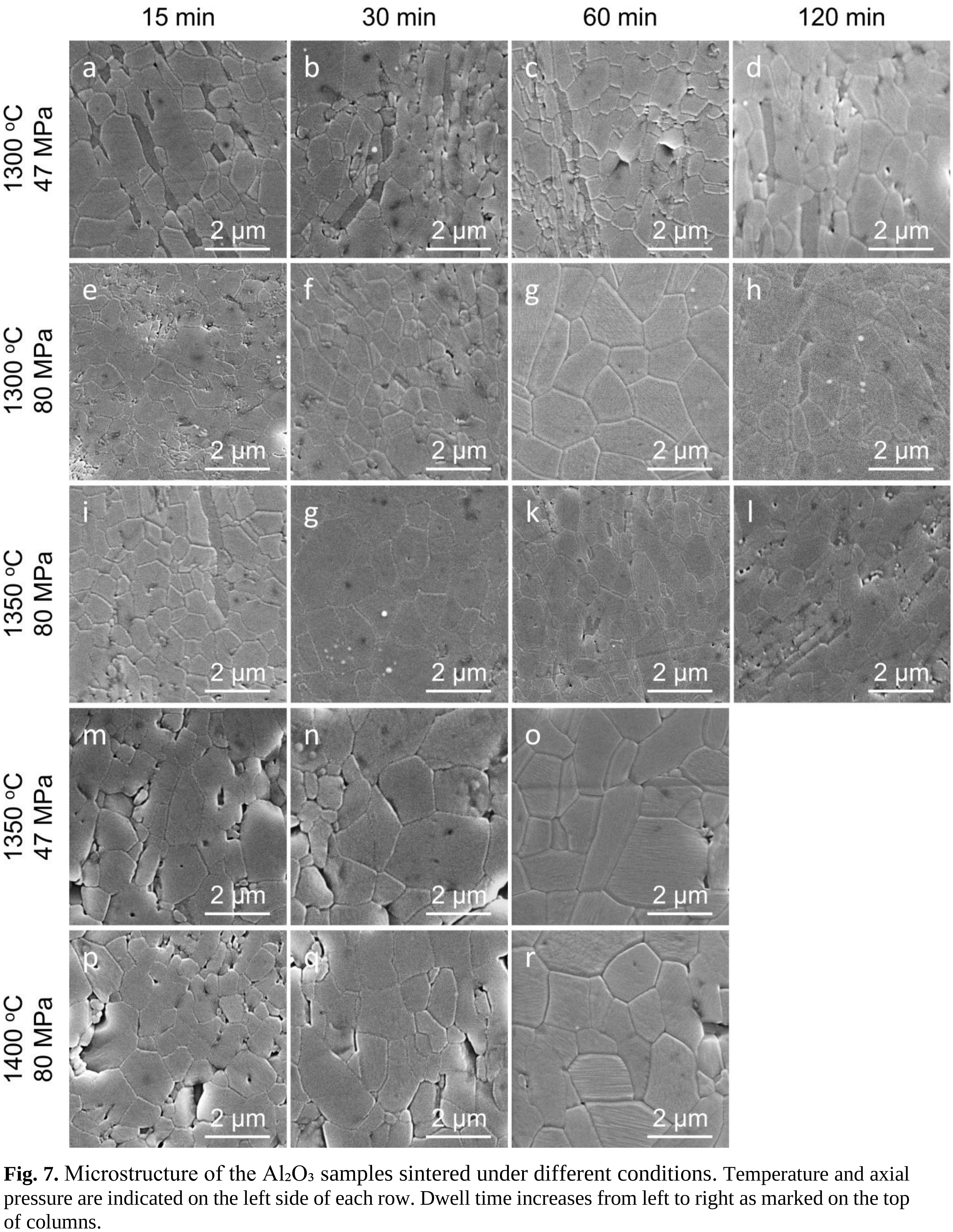

**Fig. 7.** Microstructure of the $Al_2O_3$ samples sintered under different conditions. Temperature and axial pressure are indicated on the left side of each row. Dwell time increases from left to right as marked on the top of columns.

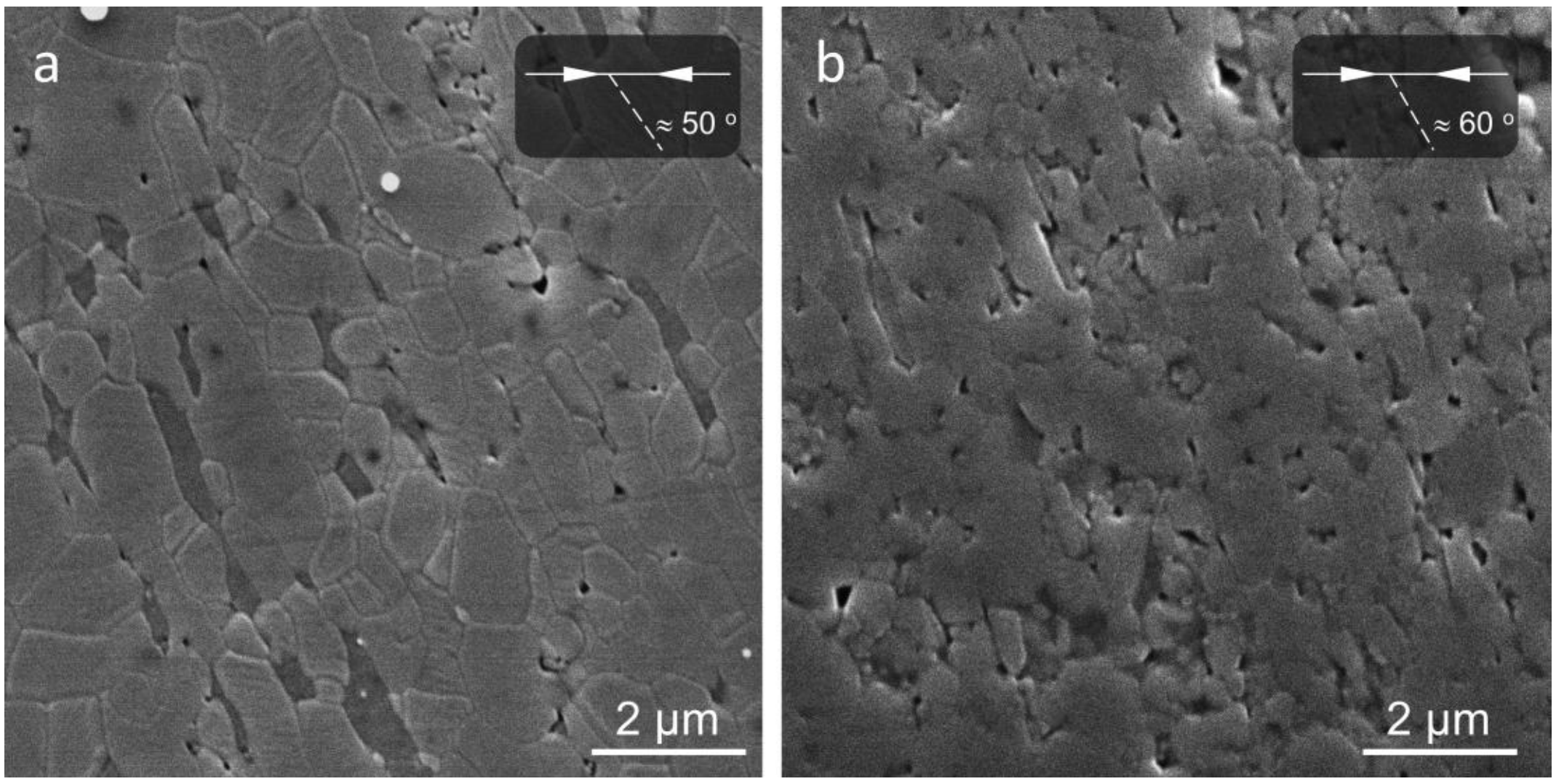

**Fig. 8.** Local alignment of elongated grain sections in the sample obtained at 1300 °C, 47 MPa, 15 min (a) and 1400 °C, 47 MPa, 15 min (b). The arrows indicate the pressing direction. The dashed line shows primary incline.

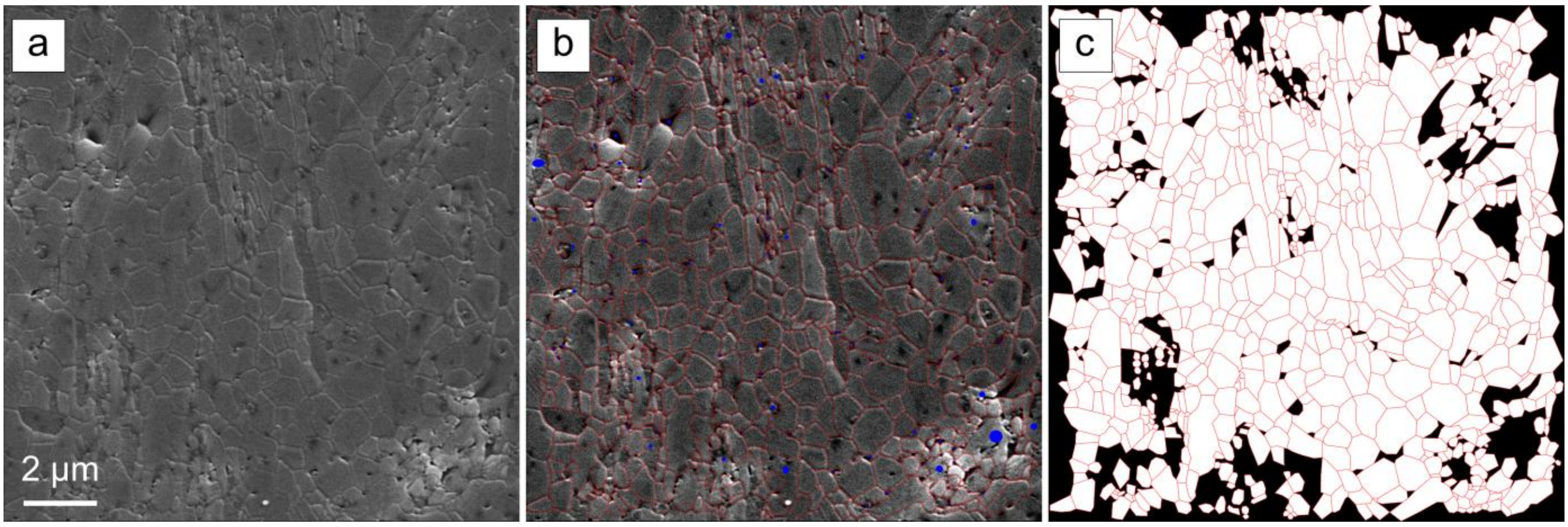

**Fig. 9.** Workflow for preparing the images of thermally etched $Al_2O_3$ for quantitative grain analysis: original micrograph of the polished and thermally etched section (a), manually traced grain boundaries overlaid on the same micrograph (b), and grain map prepared for analysis (c). Grains retained for measurement are shown in white, grain boundaries are red, blue marks represent areas for exclusion, and excluded regions are filled black.

Figure 10 shows the obtained chord-length distributions in semi-logarithmic scale. The majority of the curves has an important segment approaching to a straight line described by exponential decay

$$e^{-\beta C}, \tag{4.3}$$

with extinction coefficient $\beta$. In Fig. 10, the extinction coefficient is indicated for chord-length intervals around $C$ = 1 μm. For temperature 1400 °C and dwell time 60 min, a linear segment can not be determined and the extinction coefficient is not obtained. Probably, the non-linear behavior is observed because of insufficient chord statistics. The exponential decay of the chord-length distribution means non-correlated distribution of grain-boundary elements in space [29]. According to Fig. 10, such a behavior is typical for chords longer than approximately 0.5 μm. Small deviations from the exponential decay occasionally observed for long chords around 2 μm and greater can be explained by insufficient statistics. Short chords below 0.5 μm significantly deviate from the exponential distribution indicating a considerable correlation between grain boundary elements at such distances. At short distances, the chord-length distribution is always below the exponent, see the dashed lines in Fig. 10. The reduced number of short chords can be explained, for example, by a rather limited number of grains of size below 0.5 μm.

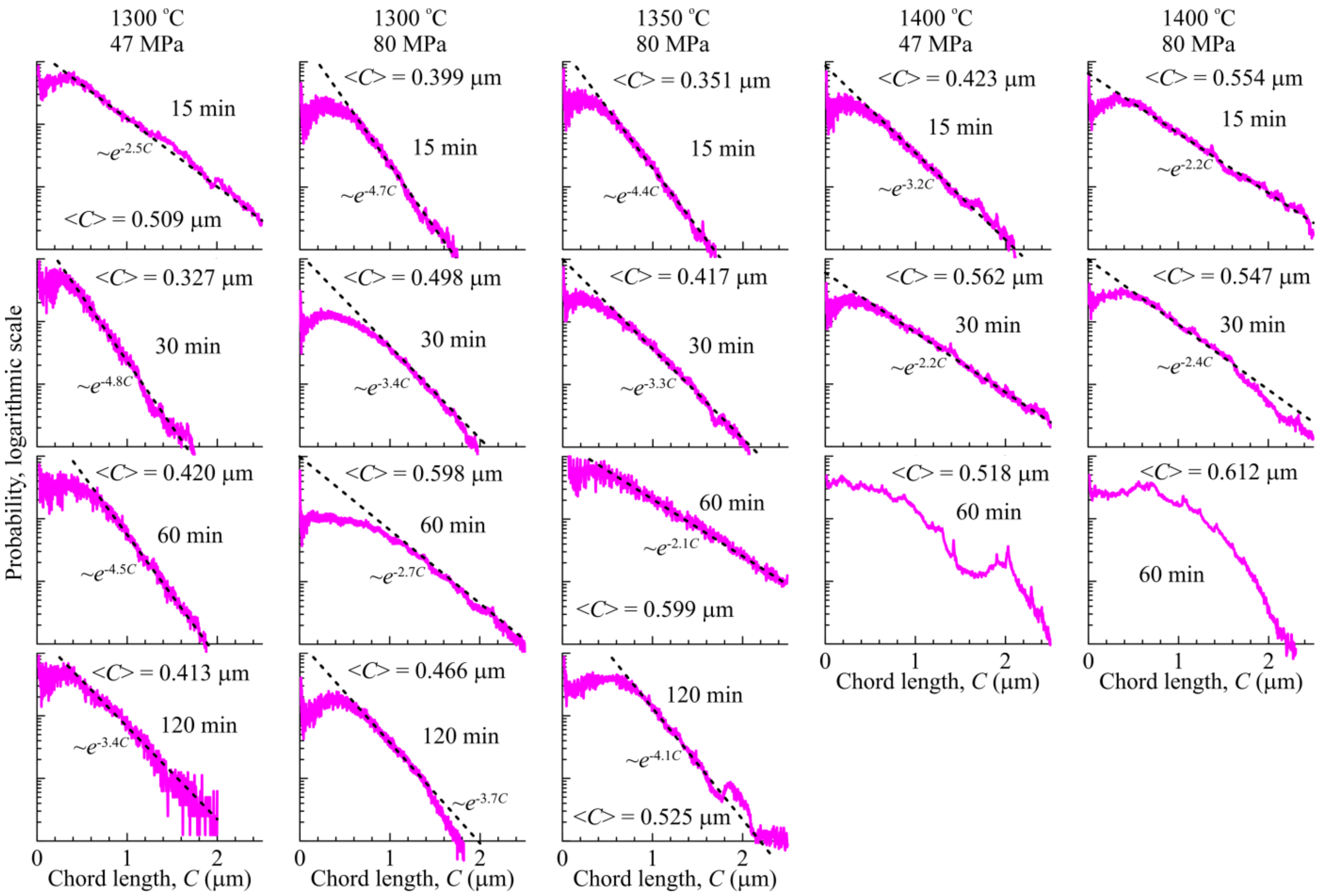

**Fig. 10.** Chord-length distribution in isothermal isobar processes. Temperature and axial pressure are indicated on the top of columns. Dwell time increases from top to bottom in the columns and indicated near distribution curves. Dashed lines are exponential fits.

Mean chord length <*C*> is calculated for the distributions and shown near the curves in Fig. 10. Table 5 summarizes the values of <*C*> and $\beta$. For ideal exponential distribution (4.3), the mean chord length is reciprocal to the extinction coefficients. Therefore, Fig. 11 compares the values of <*C*> and $1/\beta$. One can see that variations of these parameters correlate while <*C*> is systematically greater. The difference is related to the correlation in the grain boundary positions noticed above. Figures 11a and 11b show that there is no persistent trend for variation of mean chord length <*C*> with time at the temperature of 1300 °C. It seems that the value of <*C*> chaotically varies in the range of 0.3 to 0.6 μm. According to Eq. (3.24), the effective grain size is proportional to <*C*> and varies in the range approximately from 0.5 to 0.9 μm. The scatter is hardly related to grain size evolution but an initial wide grain size distribution. Figure 1b shows that initial powder particles are somewhat smaller than 0.5 μm. Thus, one can expect gran growth at dwell time below 15 min but not in the interval of time from 15 to 120 min covered by present experiments.

**Table 5.** Mean chord length, <*C*> (μm), and extinction coefficient $\beta$ (μm$^{-1}$)

| Temperature, *T* (°C) | | 1300 | 1300 | 1350 | 1400 | 1400 |
|---|---|---|---|---|---|---|
| Axial pressure, $-\sigma_{\|}$ (MPa) | | 47 | 80 | 80 | 47 | 80 |
| Dwell time, *t* (min) | 15 | <*C*> = 0.509<br>$\beta$ = 2.5 | <*C*> = 0.399<br>$\beta$ = 4.7 | <*C*> = 0.351<br>$\beta$ = 4.4 | <*C*> = 0.423<br>$\beta$ = 3.2 | <*C*> = 0.554<br>$\beta$ = 2.2 |
| | 30 | <*C*> = 0.327<br>$\beta$ = 4.8 | <*C*> = 0.498<br>$\beta$ = 3.4 | <*C*> = 0.417<br>$\beta$ = 3.3 | <*C*> = 0.562<br>$\beta$ = 2.2 | <*C*> = 0.547<br>$\beta$ = 2.4 |
| | 60 | <*C*> = 0.420<br>$\beta$ = 4.5 | <*C*> = 0.598<br>$\beta$ = 2.7 | <*C*> = 0.599<br>$\beta$ = 2.1 | <*C*> = 0.518 | <*C*> = 0.612 |
| | 120 | <*C*> = 0.413<br>$\beta$ = 3.4 | <*C*> = 0.466<br>$\beta$ = 3.7 | <*C*> = 0.525<br>$\beta$ = 4.1 | | |

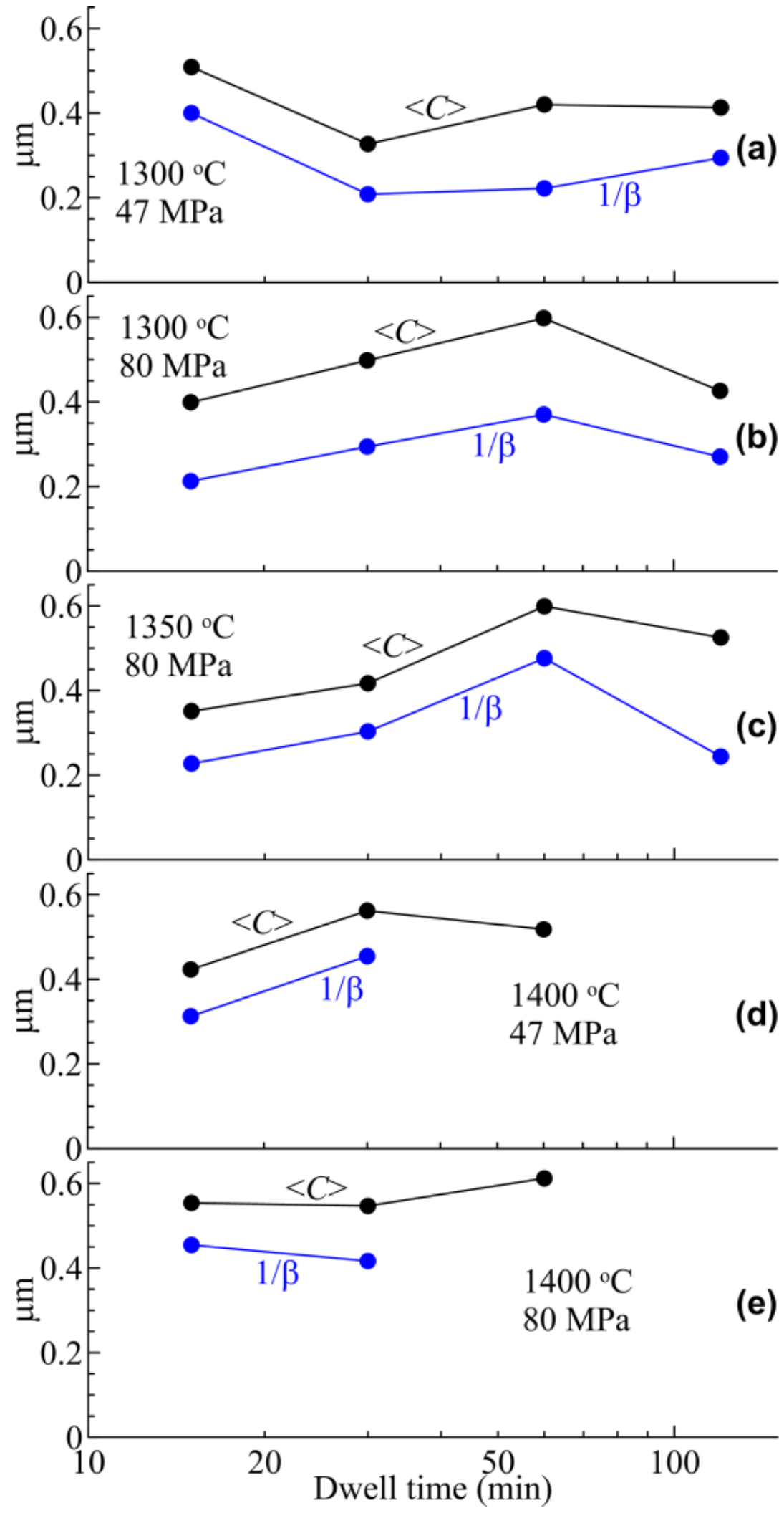

**Fig. 11.** Evolution of mean chord length <*C*> and reciprocal to extinction coefficient 1/$\beta$ in isothermal isobar hot pressing with the indicated values of temperature and axial pressure.

According to Figs. 11c-11e, the variation of <*C*> is approximately the same, from 0.4 to 0.6 mm at temperatures 1350 and 1400 °C. This corresponds to the effective grain size 0.6 to 0.9 μm. In Figs. 11d and X2c, one could notice a weak tendency of grain growth at 1400 °C. However, this tendency seems to be insignificant taking into account the scatter. Figure 12 shows the specific surface calculated by Eq. (3.23) which varies in the interval from approximately 3 to 6 μm$^{-1}$. Similarly, such a variation does not signify a systematic tendency but a scatter in the initial conditions. Besides, Figs. 12a and 12b do not reveal an influence of axial pressure. In summary, no evolution of grain size and grain boundary surface is observed in the interval of dwell time from 15 to 120 min at temperatures from 1300 to 1400 °C. There is no influence of axial pressure in the interval of 47 to 80 MPa. There are local variations with the effective grain size from approximately 0.5 to 0.9 μm and specific surface from approximately 3 to 6 $\mu m^{-1}$.

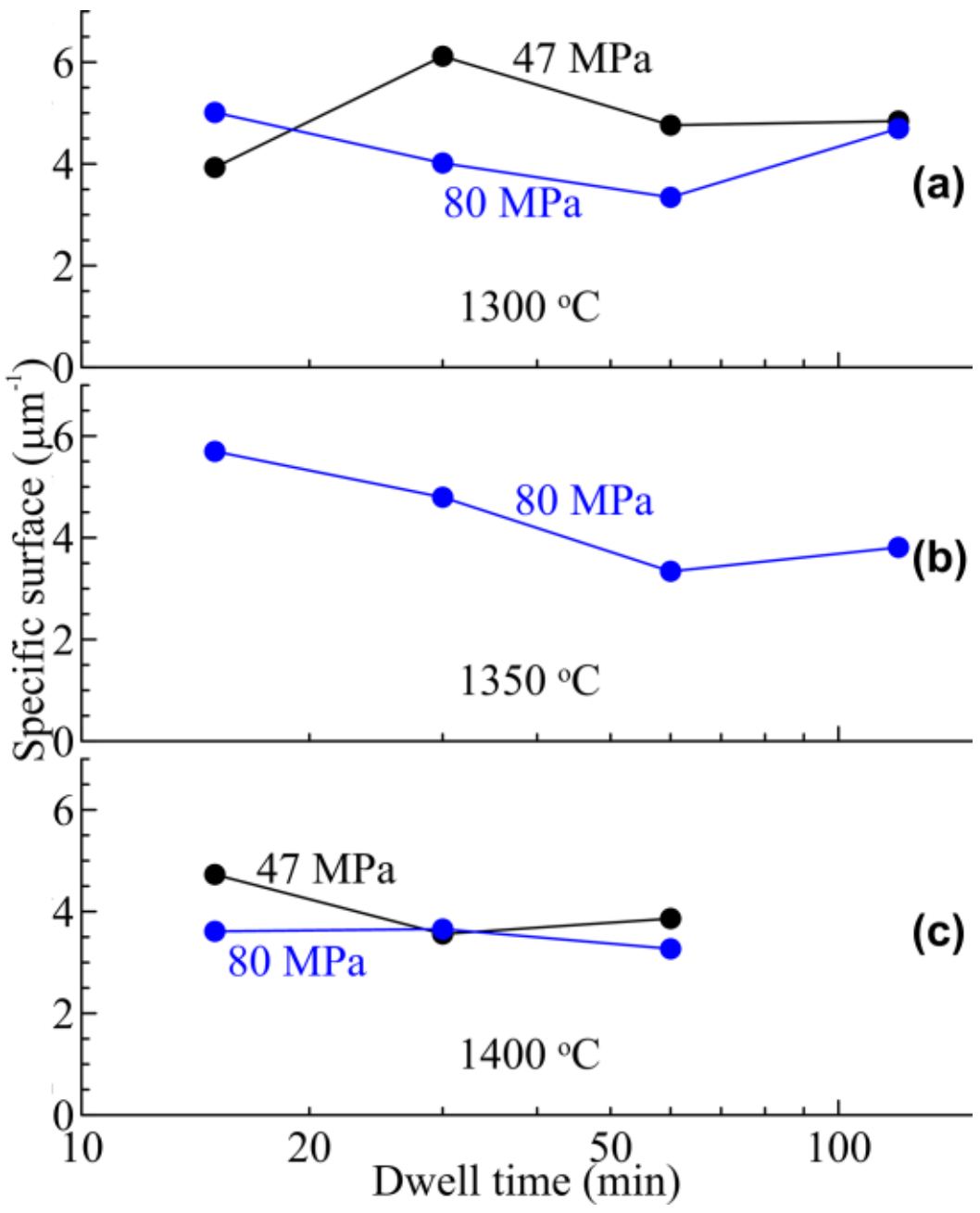

**Fig. 12.** Evolution of grain-boundary specific surface in isothermal isobar hot pressing with the indicated values of temperature and axial pressure.

## 5. Discussion

The above microstructure analysis reveals that in the studied conditions, the only essentially variable microstructure parameter is the porosity. Figure 6 shows that the densification rate strongly depends on porosity $\theta$ with approximately cubic function (4.2) for low $\theta \leq 0.2$. Factor $w$ in Eq. (4.2) is actually the extrapolation of the densification rate to the value $\theta = 1$. Therefore, below factor $w$ is referred to as the reduced densification rate. Experiments at different temperatures and pressures give different porosities. However, the results become comparable only at the same porosity value, and the reduced densification rate does characterize the densification rate at the same porosity.

### *5.1. Stress exponent*

Figure 13 shows that the densification rate significantly increases with the axial pressure. The dependence can be approximated by a power law

$$w \sim (-\sigma_{\parallel})^n , \qquad (5.1)$$

with stress exponent $n$. Lines connecting data points in Fig. 13 are the power-law approximations. The estimated stress exponent is around 3. The variations with temperature are within the experimental uncertainty.

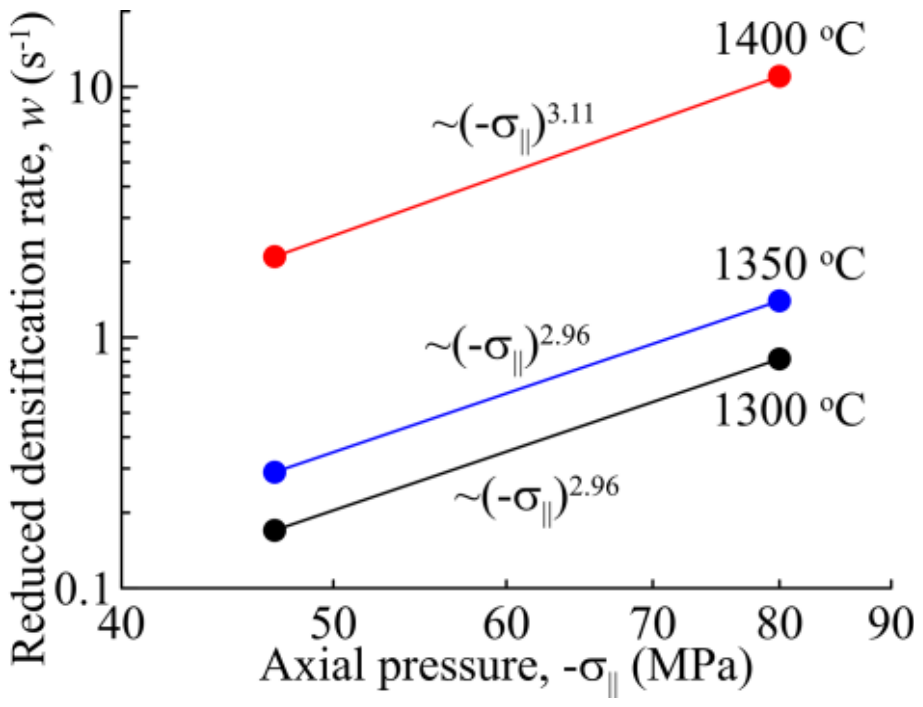

**Fig. 13.** Pressure dependence of the densification rate. Temperatures and power fits are indicated near lines connecting data points.

*5.2. Activation energy*

If the stress relaxation in alumina were controlled by an energy-barrier process, the reduced relaxation rate would follow the Arrhenius law

$$w \sim \exp\left(-\frac{Q}{RT}\right), \tag{5.2}$$

with activation energy $Q$, universal gas constant $R$, and absolute temperature $T$. In Fig. 14, the Arrhenius plot of ln($w$) versus 1/$T$ shows that the experimental points do not lie on a line in the temperature interval 1300 to 1400 °C. Thus, the relaxation is not controlled by a single energy-barrier process but rather a competition of several processes. One can introduce the apparent activation energy in temperature interval $[T_1,T_2]$ by the following equation derived from Eq. (5.2):

$$Q = R\frac{T_1 T_2 (\ln w_1 - \ln w_2)}{T_1 - T_2}. \tag{5.3}$$

The apparent activation energy is proportional to the slope of the line connecting experimental points $(w_1,T_1)$ and $(w_2,T_2)$ on the Arrhenius plot.

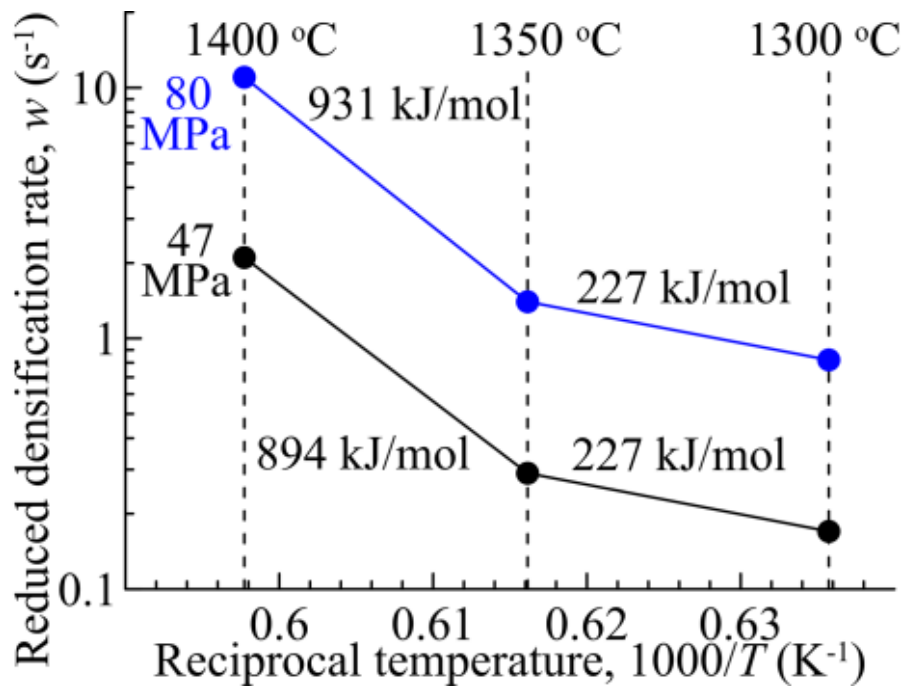

**Fig. 14.** Arrhenius plots for the densification rate at the axial pressure of: 47 MPa (black) and 80 MPa (blue). Vertical dashed lines show centigrade temperature. The value indicated near an interval connecting two experimental points is the effective activation energy in the interval.

Figure 14 shows four values of the effective activation energy estimated by Eq. (5.3) in the temperature intervals [1300,1350] °C and [1350,1400] °C at the axial pressures of 47 and 80 MPa. One can see that the effective activation energy considerably increases with temperature from approximately 230 kJ/mol at 1300-1350 °C to 900 kJ/mol at 1350-1400 °C. The experimental values of reduced densification rate $w$ obtained from Fig. 6 hardly contain more than one significant digit. Their relative uncertainty is estimated around 10%. Therefore, the relative uncertainty of the derived apparent activation energy is about 10% too. In these conditions, the dependence on the pressure seems to be not significant and rather reflects an experimental scatter.

*5.3. Mechanisms of stress relaxation*

It is well known that the phenomena of dense material creep and porous material densification at compressive load are similar. Antou *et al.* [15] reasonably used the densification rate measured at hot pressing as the value equivalent to the strain rate in creep experiments. The Maxwell-type rheology model of hot pressing developed in Section 2 of the present work is applicable to arbitrary load conditions and arbitrary porosity levels. Thus, the model underlines that creep and densification are essentially the same phenomenon of stress relaxation. The model can compare different load conditions but can not predict how the relaxation rate value depends on the temperature, pressure, and other conditions. The objective of the present section is to analyze the obtained experimental data to reveal probable stress relaxation mechanisms in the studied conditions. Such an analysis would help in understanding both the creep and densification phenomena and in developing a predictive mathematical model.

Section 5.1 concludes that in the present experiments, the effective grain size varies within a sample in the interval around 0.5-0.9 μm but grain growth is not observed in the studied interval of dwell time from 15 to 120 min. This means that the obtained experimental results concern the indicated above grain-size range and that the influence of grain size can not be revealed and the grain-size exponent can not be estimated from the present data. Figure 6 clearly shows a sharp increase of the densification rate with porosity. In the range of porosity approximately from 0.05 to 0.2, cubic dependence (4.2) is applicable. Thus, the porosity is a very important microstructure parameter in the studied conditions. The influence of porosity can not be explained by simple Nabarro-Herring or Coble diffusion mechanisms but appears reasonable in grain-boundary sliding (GBS). Indeed, stress relaxation around intergranular pores could facilitate the sliding.

The diffusion mechanisms result in the stress exponent equal to one confirmed, for example, by the results of Harmer and Brook [8]. The cited work covers the grain-size range of the present work but concerns higher porosity of 0.25 and temperature of 1475 °C. On the contrary, the stress exponent is estimated around 3 in the present work as shown in Fig. 13. Such a high value contradicts the Nabarro-Herring and Coble models. It seems that the principal difference between the conditions of the present work and those studied in Ref. [8] is the porosity level. It is known that the contribution of grain rearrangement becomes more important with decreasing the porosity [8]. Therefore, it is possible that the diffusion mechanisms control densification at the higher porosity of 0.25 studied in Ref. [8] while GBS becomes the principal one at lower porosities studied in the present work.

Cannon *et al.* [30] studied creep of full-density alumina ceramics with the porosity no greater than 0.01 in the conditions similar to those of the present work and found the stress exponent in the range from 1 to 2. Heuer *et al.* [31] explained such a high value of stress exponent by superposition of the diffusion mechanisms and dislocation slip and non-accommodated and accommodated by slip GBS. The mechanism of dislocation slip results in the stress exponent from 3 to 5 [30]. This mechanism can explain the stress exponent observed in the present work. However, dislocation glide in alumina with grain size around 1 μm requires critical resolved shear stress about 60 MPa [31] hardly attained in the present experiments. Nevertheless, such a high value can be the result of stress concentration [31]. Thus, dislocation glide is a possible mechanism of accommodation at gran-boundary sliding in the conditions of the present work. Dislocation glide is not a thermally-activated process while Fig. 14 shows a considerable temperature dependence observed in the present work. Therefore, there should be parallel thermally-activated controlling processes.

Chokshi [32] observed GBS at 30-40 MPa tensile load in polycrystalline alumina with grain size around 1 μm at 1400 °C. He studied the deformation of linear superficial marks by scanning electron microscopy and conclude that the contribution of GBS in the total strain is around 70%. Clarisse *et al.* [33] studied GBS at compressive load of 50 MPa and temperature 1350 °C by atomic force microscopy. They used full-density polycrystalline alumina samples with the grain size around 1 mm and also concluded that the contribution of GBS is around 70%. The conditions studied in the present work are very similar to those of Refs. [32] and [33]. Therefore, one might expect a considerable contribution of GBS too.

Ruano *et al.* [34] reviewed the accumulated experimental data on high-temperature deformation of fine-grained polycrystalline alumina and concluded that the considerable fraction of the results is consistent with the mechanism of GBS accommodated by slip. They supposed that the slip is controlled by dislocation climb, thus obtaining the stress exponent of two and the activation energy around that of diffusion. The reviewed experimental data could be classified into two groups, the first one with the activation energy around 600-800 kJ/mol corresponding to lattice diffusion and the second one with the activation energy around 450-500 kJ/mol corresponding to grain boundary diffusion [34]. The present experimental data indicate a considerably greater stress exponent of 3 and extremely variable apparent activation energy from 230 to 900 kJ/mol. The maximum activation energy can be consistent with a lattice-diffusion controlling process like dislocation climb while the minimum is difficult to explain by a diffusion process.

Antou *et al.* [15] observed fractured particles in submicron-grained alumina ceramics with porosity about 0.25 to 0.4 after hot pressing in the temperature range between 900 and 1100 °C. The particle fracture can be an additional mechanism for GBS accommodation. The elevated stress exponent around 2 to 3 and intermediate activation energy around 300 to 500 kJ/mol [15] can be explained by the contribution of particle fracture. Boldin *et al.* [16] reported the similar value of activation energy around 400 kJ/mol for hot pressing of alumina at 1520 °C and 70 MPa. Such activation energy is typical for grain boundary diffusion. The results of Ref. [16] can be explained by the Coble diffusion model as well as by GBS controlled by dislocation climb. It is not possible to assess the relative contribution of these mechanisms because the stress and grain-size exponents were not estimated in Ref. [16].

The conditions of the above-cited works are similar to the conditions of the present experiments. Therefore, all the considered deformation mechanisms can be important. To understand the interplay of the mechanisms, the following experimental findings of the present work are principal: (1) the stress relaxation rate strongly increases with the porosity, (2) the stress exponent is approximately 3, and (3) the apparent activation energy can be considerably lower than that of grain-boundary diffusion. The first statement can be explained by the commonly accepted superposition of diffusion Nabarro-Herring and Coble creep with GBS accommodated by slip because GBS implies rearrangement of grains which is reasonably facilitated by pores. However, the frequently accepted GBS accommodation controlled by dislocation climb results in the stress exponent of two and the activation energy equal to that of diffusion and cannot explain statements (2) and (3). There should be additional mechanisms of GBS accommodation and deformation as particle fracture or dislocation glide. Such processes are not thermally activated and ensure the decrease of the apparent activation energy. A threshold stress should be attained to start the fracture and the dislocation glide. This results in a highly nonlinear behavior increasing the stress exponent.

Dislocation slip could explain the observed high stress exponent and low activation energy. However, a rather high threshold value of shear stress is necessary, the so-called critical resolved shear stress (CRSS). In the present work, the axial pressure attains 80 MPa while CRSS is estimated around 60 MPa in the similar conditions [31]. The issue is that at uniaxial hot pressing, the stress state is about isotropic compression and hardly results in a shear stress above CRSS. Indeed, Eqs. (2.18) and (2.27) relate the values of pressure $p$ and axial $\sigma_{\parallel}$ and lateral $\sigma_{\perp}$ normal stresses at uniaxial compression. The maximum shear stress equals half the difference between maximum $\sigma_{\perp}$ and minimum $\sigma_{\parallel}$ principal stresses,

$$\tau_{\max} = \frac{\sigma_{\perp} - \sigma_{\parallel}}{2} = -\frac{3}{4}\sigma_{\parallel}\frac{\theta(1-\theta)^{2/3}}{1+\theta(1-\theta)^{2/3}} . \tag{5.4}$$

At axial pressure $-\sigma_{\parallel} = 80$ MPa and porosity $\theta = 0.1$, the maximum shear stress is as low as $\tau_{\max} = 5$ MPa. Such a value seems to be insufficient for dislocation sleep. Nevertheless, stresses can concentrate around pores where the conditions for dislocation sleep could satisfy. The next section aims to estimate stresses around pores.

*5.4. Stress concentration around pores*

At uniaxial compression, the stress state can be considered as a superposition of isotropic compression with pressure $p$ and pure shear with maximum shear stress $\tau_{\max}$. According to the above estimates for the present work, the shear stress around 5 MPa is much less that the pressure around 80 MPa. Therefore, it is reasonable to start with the principal isotropic component. The model problem of a spherical pore in an infinite elastic medium is well known. Landau and Lifshitz [19] analyzed the stress field at isotropic compression. The equivalent problem is stated in Appendix A for a spherical layer. When external radius $R_2$ tends to infinity, Eqs. (A8) and (A9) result in the classical distribution for the normal radial stress [19],

$$\sigma_{\mathrm{rr}} = p\left[\left(\frac{R_1}{r}\right)^3 - 1\right]. \tag{5.5}$$

Equation (A5) gives normal tangential stress

$$\sigma_{\tau\tau} = -p\left[\frac{1}{2}\left(\frac{R_1}{r}\right)^3 + 1\right]. \tag{5.6}$$

Maximum shear stress

$$\frac{\sigma_{\mathrm{rr}} - \sigma_{\tau\tau}}{2} = \frac{3}{4} p \left( \frac{R_1}{r} \right)^3, \tag{5.7}$$

attains the maximum value of $3p/4$ on the pore surface, $r = R_1$. At isotropic compression with the pressure of 80 MPa, stress concentration around a spherical pore may result the shear stress as high as 60 MPa according to the above estimate. Such a level of shear stress is around CRSS and makes realistic the contribution of dislocation slip.

Timoshenko and Goodier [35] found the stress field around a spherical pore in infinite elastic medium at pure shear and derived stress-concentration factor

$$\frac{15(1-\nu)}{7-5\nu}, \tag{5.8}$$

where $\nu$ is the Poisson's ratio. The factor is around two for the value of $\nu = 0.25$ typical for alumina. This means that at the pure shear of 5 MPa, the maximum shear stress around a spherical pore attains approximately 10 MPa. Vakaeva *et al.* [36] numerically calculated the stress field around pores of rounded triangular cross section typically arising between grains at their triple junction and reported the stress-concentration factor up to 6. Thus, the pure shear of 5 MPa may result the shear stress of 30 MPa near the pore. In summary, even the relatively weak pure shear component may result in considerable local shear stresses concentrated around pores. If both the isotropic compression and pure shear components of stress field may give local shear stresses comparable to CRSS separately, their superposition at hot pressing is likely to increase the maximum shear stress and to widen the domains of stress concentration around pores where CRSS is attained. Thus, the combination of viscoelastic deformation with local plastic flow seems to be realistic in hot pressing.

*5.5. Nonlinear viscoelastic model*

The accumulated experimental data show that the high-temperature strain rate of alumina can be a strongly nonlinear function of applied stresses resulting a stress exponent considerably greater than one and the present work confirms this statement once more. The straightforward way to explain such a nonlinearity is to assume that the relaxation rate tensor $\mathbf{W}$ depends on stress tensor $\boldsymbol{\sigma}$ in the viscoelastic constitutive relation (2.12). However, the Clausius-Duhem inequality (2.13) no more reduces to the non-negativity of a quadratic form and requires additional analysis. Consider quadratic form

$$\mathbf{W}(\mathbf{s}) : \boldsymbol{\sigma} : \boldsymbol{\sigma}, \tag{5.9}$$

where tensor $\mathbf{s}$ is a parameter. If the form is non-negative at any value of $\mathbf{s}$, substitution $\mathbf{s} = \boldsymbol{\sigma}$ results in a non-negative value. Thus, non-negativity of quadratic form (5.9) is sufficient to satisfy the Clausius-Duhem inequality

$$\mathbf{W}(\boldsymbol{\sigma}) : \boldsymbol{\sigma} : \boldsymbol{\sigma} \geq 0. \tag{5.10}$$

In isotropic media where the relaxation rate tensor (2.15) is defined by two scalars, the rates of bulk $\omega_b$ and shear $\omega_s$ relaxation, the non-negativity of functions $\omega_b(\boldsymbol{\sigma})$ and $\omega_s(\boldsymbol{\sigma})$ is sufficient for the Clausius-Duhem inequality. In isotropic media, the invariance about rotations requires that these functions are functions of scalar invariants of stress tensor $\boldsymbol{\sigma}$. It is reasonable to assume that the shear relaxation rate depends on the maximum shear stress $\tau_{\max}$. In particular, power law for stress relaxation rate at full density

$$\omega_{s0} = \Omega_n \tau_{\max}^{n-1}, \tag{5.11}$$

where $\Omega_n$ is a constant independent of $\boldsymbol{\sigma}$, gives stress exponent of $n$ at pure-shear stress state. In Eq. (5.11), the exponent is one less than the pressure exponent because in constitutive relation (2.12), the shear stress relaxation rate is multiplied by the deviatoric component of $\boldsymbol{\sigma}$ which is proportional to $\tau_{\max}$.

Appendix A has shown that the model problem of spherical layer is useful to find functional relation between shear relaxation rate of dense material and bulk relaxation rate of porous material. Appendix B solves such a problem for nonlinear viscoelastic medium with shear relaxation rate (5.11) and obtains power-law densification rate (B8) with the same stress exponent $n$. The difference is that the driving force for densification is not the shear stress but the pressure. Equation (B9) relates the nonlinear bulk and shear relaxation rates. If factor $\Omega_n$ defining the shear relaxation rate is non-negative, both shear (5.11) and bulk (B9) relaxation rates are non-negative. Thus, the Clausius-Duhem inequality is satisfied.

The problems of isostatic and uniaxial pressing stated in Section 2 can be solved for nonlinear shear (5.11) and bulk (B9) relaxation rates resulting the following differential equation of densification:

$$\frac{(1-\theta^{1/n})^n}{\theta(1-\theta)}\dot{\theta} = -\frac{3^n}{2^{2n-1}n^n}\frac{\Omega_n}{\mu_0}p^n, \tag{5.12}$$

for isostatic pressing and

$$\left\{\frac{1-\theta^{1/n}}{\left[\theta(1-\theta)\right]^{1/n}} + \frac{(1-\theta)^{\frac{2}{3n}}}{n}\right\}^n \dot{\theta} = -\frac{3^n}{2^{2n-1}n^n}\frac{\Omega_n}{\mu_0}(-\sigma_{\|})^n, \tag{5.13}$$

for uniaxial pressing. The ratio of axial stress to pressure at uniaxial pressing becomes

$$-\frac{\sigma_{\|}}{p} = 1 + \frac{\theta^{1/n}(1-\theta)^{\frac{5}{3n}}}{n(1-\theta^{1/n})}. \tag{5.14}$$

Equations (5.12)-(5.14) generalize Eqs. (2.25)-(2.27), respectively. Substitution of pressure exponent $n = 1$ into Eqs. (5.12)-(5.14) reduces them to Eqs. (2.25)-(2.27).

The studied nonlinear viscoelastic model shows that the stress exponent is the same in pure shear of a full-density material and isotropic compression of a porous material. The model formally explains non-linear strain rate-load dependencies in hot pressing and creep experiments. According to the model, the observed elevated values of stress exponent of 3 and higher are not specific for the hot-pressing load conditions and should be observed in creep experiments too. Shear (5.11) at bulk (B9) relaxation rates are proportional to the same factor $\Omega_n$. The factor is independent of stress state but responsible for other conditions including temperature. Thus, the model assumes similar temperature dependencies for the shear and bulk relaxation rates. If shear and bulk relaxation are thermally activated processes, they should have the same activation energy. In view of the discussion in Section 5.3, the nonlinear viscoelastic model is hardly useful to explain the observed reduced apparent activation energies.

According to Fig. 5, the experimentally measured porosity decreases as a power of dwell time. At low porosity $\theta$, densification equation (5.13) becomes

$$\frac{\dot{\theta}}{\theta} = -\frac{3^n}{2^{2n-1}n^n}\frac{\Omega_n}{\mu_0}(-\sigma_{\|})^n . \qquad (5.15)$$

If the right-hand side of this differential equation is a constant, it results in an exponential decay of porosity. To explain the observed time exponent of -1/2, one should assume that factor $\Omega_n$ depends on porosity as $\theta$ squared. This contradicts the logics of the model because factor $\Omega_n$ is introduced by Eq. (5.11) for a full-density material. The same drawback characterizes the linear model corresponding to $n = 1$. In summary, the nonlinearity introduced into the viscoelastic model is useful to describe the observed elevated values of stress exponent but not advantageous in explaining the reduced apparent activation energies and the dependence of the relaxation rate on porosity.

*5.6. Model of combined viscoelastic and plastic flow*

Section 5.4 concludes that in the conditions of present experiments, stress concentration around pores may result in the maximum shear stress $\tau_{max}$ attaining critical resolved shear stress (CRSS) $\tau_c$. Thus, domains of plastic flow may form around pores. Appendix C solves the problem of spherical layer with coexisting domains of viscoelastic and plastic flow to obtain the relation between shear and bulk relaxation rates. In the viscoelastic domain where $\tau_{max} < \tau_c$, shear relaxation rate is specified by nonlinear equation (5.11). In the domain of plastic flow where the maximum shear stress attains CRSS, the shear relaxation rate generally increases to maintain equality $\tau_{max} = \tau_c$.

Densification rate (C8) and bulk relaxation rate (C9) are obtained. If factor $\Omega_n$ defining the shear relaxation rate is non-negative, shear relaxation rate in the viscoelastic (5.11) and plastic (C11) domains and bulk relaxation rate (C9) are non-negative. Section 5.5 states that the non-negativity of these rates is sufficient to satisfy the Clausius-Duhem inequality.

The problems of isostatic and uniaxial pressing stated in Section 2 can be solved for nonlinear shear, (5.11) and (C11), and bulk (C9) relaxation rates. At isostatic pressing, densification rate

$$\frac{\dot{\theta}}{1-\theta} = -\omega_c \phi , \qquad (5.16)$$

where characteristic frequency $\omega_c$ and dimensionless parameter $\phi$ are defined as follows:

$$\omega_c = \frac{2\Omega_n \tau_c^n}{\mu_0} , \qquad \phi = \left(\frac{R_c}{R_2}\right)^3 . \qquad (5.17)$$

According to the model of spherical layer, parameter $\phi$ is the sum of porosity and the volume fraction of the plastic domain. Complementary value $1 - \phi$ is the volume fraction of the viscoelastic domain. The nonlinear function $\phi(\theta)$ is defined by transcendent equation (C6) written as

$$\ln\phi - \ln\theta + n(1-\phi^{1/n}) = \frac{3p}{4\tau_c} . \qquad (5.18)$$

When the plastic domain vanishes, $\phi = \theta$ and the exclusion of parameter $\tau_c$ from Eqs. (5.16) and (5.18) results in Eq. (5.12) derived above for the nonlinear viscoelastic model.

At uniaxial pressing, the same differential equation (5.16) is applicable but Eq. (5.18) becomes useless to find function $\phi(\theta)$ because it contains unknown pressure $p$. The ratio of axial stress to pressure becomes

$$-\frac{\sigma_{\|}}{p}=1+\frac{\phi^{1/n}(1-\theta)^{\frac{5}{3n}}}{\ln\phi-\ln\theta+n(1-\phi^{1/n})}. \tag{5.19}$$

Substitution of Eq. (5.19) into Eq. (5.18) gives the following transcendent equation defining function $\phi(\theta)$ at the given value of axial stress $\sigma_{\|}$:

$$\ln\phi-\ln\theta+n(1-\phi^{1/n})+\phi^{1/n}(1-\theta)^{\frac{5}{3n}}=-\frac{3\sigma_{\|}}{4\tau_{c}}. \tag{5.20}$$

When the plastic domain vanishes, $\phi = \theta$, Eq. (5.19) reduces to Eq. (5.14), and the exclusion of parameter $\tau_c$ from Eqs. (5.16) and (5.20) results in Eq. (5.13) for the nonlinear viscoelastic model.

Equation (5.16) shows that densification rate (4.1) is proportional to parameter $\phi$. At isostatic pressing, $\phi$ is a function of porosity $\theta$ and pressure $p$ specified by Eq. (5.18) while at uniaxial pressing, $\phi$ is a function of porosity $\theta$ and axial pressure $-\sigma_{\|}$ specified by Eq. (5.20). At constant porosity, the functions are nonlinear and do not reduce to a power law. However, one can define apparent stress exponent of densification in isostatic $n_i$ and uniaxial $n_u$ pressing as

$$n_{i}=p\frac{d\phi}{dp}, \qquad n_{u}=\sigma_{\|}\frac{d\phi}{d\sigma_{\|}}, \tag{5.21}$$

where differentiation assumes constant porosity. Equations (5.18) and (5.20) result, respectively

$$n_{i}=\frac{3p}{4\tau_{c}}\frac{1}{1-\phi^{1/n}}, \qquad n_{u}=-\frac{3\sigma_{\|}}{4\tau_{c}}\frac{1}{1-\left[1-\frac{1}{n}(1-\theta)^{\frac{5}{3n}}\right]\phi^{1/n}}. \tag{5.22}$$

The both expressions result pressure exponent at pure shear $n$ when the plastic domains vanishes but they sharply increase with pressure $p$ or axial pressure $-\sigma_{\|}$, respectively, above plastic flow threshold (C10). Thus, the apparent stress exponent of densification can be considerably greater than the stress exponent at pure shear if plastic flow develops.

Analysis of Eqs. (5.18) and (5.20) indicates that at low porosity $\theta \ll 1$, parameter $\phi$ becomes approximately proportional to $\theta$. Assuming the proportionality, Eq. (5.16) results in exponential decay of $\theta$ with time like the considered above models without plastic flow which contradicts the experimentally observed power-law decay. The temperature dependence of densification rate (5.16) is determined by characteristic frequency $\omega_c$ given by Eq. (5.17). According to the Arrhenius law (5.2), the apparent activation energy is defined as

$$Q=RT^{2}\frac{d\ln\omega_{c}}{dT}=RT^{2}\frac{d\ln\Omega_{n}}{dT}+nRT^{2}\frac{d\ln\tau_{c}}{dT}-RT^{2}\frac{d\ln\mu_{0}}{dT}. \tag{5.23}$$

The first term in the right-hand side of Eq. (5.23) corresponds to the activation energy at pure shear. The second term is negative if CRSS $\tau_c$ reasonably decreases with temperature. The influence of the second term increases with pressure exponent at pure shear $n$ and can explain the experimentally observed low values of apparent activation energy. In summary, the model of combined viscoelastic and plastic flow is not advantageous in explaining the experimentally observed dependence of the densification rate on porosity but useful to describe the elevated values of stress exponent and reduced values of apparent activation energy at densification.

## 6. Conclusion

Densification kinetics of α-alumina polydisperse submicron powder at uniaxial hot pressing is studied in isothermal isobar experiments in the ranges of temperature 1300-1400 °C and pressure 47-80 MPa.

- Metallographic chord analysis of sintered samples does not reveal grain growth in the interval of dwell times 15 to 120 min. Local microstructure variations within a sample are observed with the effective grain size from approximately 0.5 to 0.9 μm and specific surface from approximately 3 to 6 $\mu m^{-1}$. The principal structural parameter in the present experiments is porosity.
- When the porosity becomes below 0.2, it approximately decays as dwell time power -1/2.
- The densification rate is approximately proportional to the porosity cubed at the porosity below 0.2. Such a strong correlation can be related to the grain-boundary sliding (GBS) mechanism.
- The stress exponent of densification is around 3 and independent of temperature. Such a high value can be explained by combination of different mechanisms of GBS accommodation including slip controlled by dislocation climb and dislocation slip.
- The apparent activation energy of densification is independent of pressure but increases with temperature from approximately 230 kJ/mol at 1300-1350 °C to 900 kJ/mol at 1350-1400 °C. The value of 230 kJ/mol is well below the diffusion activation energy and can be due to participation of non-thermally activated processes as dislocation glide.
- A theoretical approach based on the Maxwell viscoelastic model is developed to describe high-temperature material deformation at arbitrary load including the conditions of hot-pressing and creep experiments. Plastic flow in the domains of stress concentration around pores can explain the increased stress exponent and the reduced apparent activation energy at densification.

## CRediT authorship contribution statement

**Andrey V. Gusarov:** Writing – original draft, Methodology, Investigation. **Pavel A. Podrabinnik:** Writing – original draft, Investigation. **Nestor Washington Solís Pinargote:** Writing – review & editing, Methodology, Investigation. **Ekaterina V. Kuznetsova:** Writing – original draft, Investigation. **Yuri O. Pristinskiy:** Investigation. **Tatiana V. Tarasova:** Writing – review & editing, Methodology. **Sergey N. Grigoriev:** Writing – review & editing, Conceptualization.

## Declaration of competing interest

The authors declare that they have no known competing financial interests or personal relationships that could have appeared to influence the work reported in this paper.

## Data availability

Data will be made available on request.

## Acknowledgments

This work was supported by the Russian Science Foundation (Grant Agreement No. 26-19-00750, https://rscf.ru/project/26-19-00750/).

## Appendix A. Isotropic compression of a viscoelastic spherical layer

To estimate function $\omega_b(\rho)$, consider isotropic deformation of a hollow sphere with external radius $R_2$ and the radius of a concentric spherical cavity $R_1 < R_2$. In the interval of distances $r$ from the center between $R_1$ and $R_2$, the medium is isotropic with full density $\rho_0$, elastic constants $K_0$ and $\mu_0$, and relaxation rates $\omega_{b0}$ = 0 and $\omega_{s0}$. Assuming spherical symmetry, the stress and strain rate tensors are specified by their normal components in the radial direction designated by index $r$ and in a perpendicular direction designated by index $\tau$. Constitutive relation (2.12) becomes

$$\begin{cases} (\dot{\sigma}_{rr} + 2\dot{\sigma}_{\tau\tau}) = 3K_0(e_{rr} + 2e_{\tau\tau}) \\ (\dot{\sigma}_{rr} - \dot{\sigma}_{\tau\tau}) + 2\omega_{s0}(\sigma_{rr} - \sigma_{\tau\tau}) = 2\mu_0(e_{rr} - e_{\tau\tau}) \end{cases}. \qquad (A1)$$

Assume that elastic deformations are small. This means that stresses are much lower than the elastic moduli. Therefore, material derivatives of stress components in the left-hand side of Eq. (A1) can be neglected compared to the right-hand side. Thus, Eq. (A1) becomes

$$\begin{cases} e_{rr} + 2e_{\tau\tau} = 0 \\ \omega_{s0}(\sigma_{rr} - \sigma_{\tau\tau}) = \mu_0(e_{rr} - e_{\tau\tau}) \end{cases}. \qquad (A2)$$

The upper line of Eq. (A2) means incompressibility. The bottom line describes relaxation of shear stresses.

Radial distributions of stresses and strain rates can be found from the conservation laws of mass and momentum, respectively,

$$\nabla \cdot \mathbf{u} = 0, \quad \nabla \cdot \boldsymbol{\sigma} = 0. \qquad (A3)$$

The left Eq. (A3) is equivalent to the upper line of Eq. (A2) where strain rate components and velocity divergence are related to the radial velocity distribution $u_r$ as follows:

$$e_{\mathrm{rr}}=\frac{\mathrm{d}u_{\mathrm{r}}}{\mathrm{d}r}\,,\qquad e_{\tau\tau}=\frac{u_{\mathrm{r}}}{r}\,,\qquad \nabla\cdot\mathbf{u}=\frac{1}{r^2}\frac{\mathrm{d}(r^2u_{\mathrm{r}})}{\mathrm{d}r}\,. \tag{A4}$$

In the considered case of spherical symmetry, the right Eq. (A3) becomes

$$\frac{\mathrm{d}\sigma_{\mathrm{rr}}}{\mathrm{d}r}+\frac{2}{r}(\sigma_{\mathrm{rr}}-\sigma_{\tau\tau})=0\,. \tag{A5}$$

The general solution of the continuity equation (upper line of Eq. (A2) or left Eq. (A3)) is

$$u_{\mathrm{r}}=\frac{A}{r^2}\,,\qquad e_{\mathrm{rr}}=-\frac{2A}{r^3}\,,\qquad e_{\tau\tau}=\frac{A}{r^3}\,, \tag{A6}$$

where $A$ is an arbitrary constant. Substitution of the constitutive relation given by the bottom line of Eq. (A2) into the momentum equation (A5) and accounting for Eq. (A6) results in the following equation for the radial normal stress:

$$\frac{\mathrm{d}\sigma_{\mathrm{rr}}}{\mathrm{d}r}=\frac{\mu_0}{\omega_{\mathrm{s0}}}\frac{6A}{r^4}\,, \tag{A7}$$

with the general solution

$$\sigma_{\mathrm{rr}}=-\frac{2\mu_0}{\omega_{\mathrm{s0}}}\frac{A}{r^3}+B\,, \tag{A8}$$

where $B$ is another arbitrary constant.

Let the pressure in the cavity of the considered hollow sphere be zero and an external pressure $p$ is applied to the external surface of the sphere. Thus, boundary conditions for Eq. (A8) and the resulting values of constants $A$ and $B$ are

$$\sigma_{\mathrm{rr}}\big|_{r=R_1}=0\,,\qquad \sigma_{\mathrm{rr}}\big|_{r=R_2}=-p\,,\qquad A=\frac{\omega_{\mathrm{s0}}p}{2\mu_0}\frac{R_1^3R_2^3}{R_1^3-R_2^3}\,,\qquad B=\frac{pR_2^3}{R_1^3-R_2^3}\,. \tag{A9}$$

The obtained solution results in the deformation rate of the hollow sphere

$$\frac{\mathrm{d}R_2}{\mathrm{d}t}=u_{\mathrm{r}}\big|_{r=R_2}=\frac{\omega_{\mathrm{s0}}p}{2\mu_0}\frac{R_1^3R_2}{R_1^3-R_2^3}\,. \tag{A10}$$

Note that the difference of the external and internal radii cubed is proportional to the volume occupied by the medium in the hollow sphere and is therefore a constant designated below as $R_0^3$. The effective density of the hollow sphere $\rho$ can be defined as follows accounting for the central cavity and surrounding full-density spherical layer:

$$\frac{\rho}{\rho_0}=\frac{R_0^3}{R_2^3}\,,\qquad R_0^3=R_2^3-R_1^3\,. \tag{A11}$$

Equation (A10) is rewritten in terms of the introduced effective density as

$$\frac{\mathrm{d}\rho}{\mathrm{d}t}=\frac{3\omega_{\mathrm{s0}}p}{2\mu_0}(\rho_0-\rho)\,. \tag{A12}$$

This equation is similar to Eq. (2.17) describing isostatic pressing of a porous medium because the first term in the left-hand side of Eq. (2.17) vanishes at a constant pressure $p$ and material derivative is equivalent to time derivative in the given conditions. Comparison of Eqs. (2.17) and (A12) indicates that the same densification rate is attained when the bulk relaxation rate is

$$\omega_{\mathrm{b}} = \omega_{\mathrm{s}0} \frac{K}{2\mu_0} \left( \frac{\rho_0}{\rho} - 1 \right). \tag{A13}$$

The obtained expression ensures zero bulk relaxation at full density $\rho_0$.

## Appendix B. Nonlinear viscoelastic model for a spherical layer

Isotropic compression of a hollow sphere is considered. The geometry is the same as in Appendix A. The difference is that the shear stress relaxation rate is not a constant but a function of stresses specified by Eq. (5.11). The objective is to obtain a model relation between the shear and bulk relaxation rates.

The maximum shear stress equals the half-difference of the maximum and minimum principal stresses. In the present problem,

$$\tau_{\max} = \frac{\sigma_{\mathrm{rr}} - \sigma_{\tau\tau}}{2}. \tag{B1}$$

The lower line of Eq. (A2) becomes

$$2\Omega_n (\tau_{\max})^n = \mu_0 (e_{\mathrm{rr}} - e_{\tau\tau}) = -3\mu_0 \frac{A}{r^3}. \tag{B2}$$

The non-linearity of the constitutive relation does not touch continuity equation (A4). Therefore, its general solution (A6) is substituted into the right-hand side of Eq. (B2). Equation (B2) results in the general solution for the maximum shear stress

$$\tau_{\max} = \left( -\frac{3\mu_0 A}{2\Omega_n} \right)^{1/n} \frac{1}{r^{3/n}}. \tag{B3}$$

Momentum equation (A5) is written as

$$\frac{\mathrm{d}\sigma_{\mathrm{rr}}}{\mathrm{d}r} + \frac{4}{r} \tau_{\max} = 0. \tag{B4}$$

The following general solution for the radial normal stress is obtained from Eqs. (B3) and (B4):

$$\sigma_{\mathrm{rr}} = \frac{4n}{3} \left( -\frac{3\mu_0 A}{2\Omega_n} \right)^{1/n} \frac{1}{r^{3/n}} + B, \tag{B5}$$

with arbitrary constants $A$ and $B$. Boundary conditions (A9) are satisfied by the following values of the constants:

$$A = -\frac{3^{n-1}}{2^{2n-1} n^n} \frac{\Omega_n p^n}{\mu_0} \frac{R_1^3 R_2^3}{(R_2^{3/n} - R_1^{3/n})^n}, \qquad B = -p \frac{R_2^{3/n}}{R_2^{3/n} - R_1^{3/n}}. \tag{B6}$$

The deformation rate of the hollow sphere is

$$\frac{\mathrm{d}R_2}{\mathrm{d}t} = u_{\mathrm{r}} \big|_{r=R_2} = \frac{A}{R_2^2} = -\frac{3^{n-1}}{2^{2n-1} n^n} \frac{\Omega_n p^n}{\mu_0} \frac{R_1^3 R_2}{(R_2^{3/n} - R_1^{3/n})^n}. \tag{B7}$$

If apparent density of the hollow sphere $\rho$ is defined by Eq. (A11), densification rate is

$$\frac{\dot{\rho}}{\rho} = -\frac{3}{R_2} \frac{\mathrm{d}R_2}{\mathrm{d}t} = \frac{3^n}{2^{2n-1} n^n} \frac{\Omega_n p^n}{\mu_0} \frac{R_1^3}{(R_2^{3/n} - R_1^{3/n})^n}. \tag{B8}$$

Comparison of Eqs. (2.17) and (B8) indicates that the same densification rate is attained when the bulk relaxation rate is

$$\omega_{\rm b} = \frac{3^{n-1}}{2^{2n-1} n^n} \frac{K}{\mu_0} \Omega_n p^{n-1} \frac{\theta}{(1-\theta^{1/n})^n} . \tag{B9}$$

Here, radii $R_1$ and $R_2$ are excluded using the apparent porosity of the hollow sphere

$$\theta = \left( \frac{R_1}{R_2} \right)^3 . \tag{B10}$$

Equation (B9) reduces to Eq. (A13) at pressure exponent $n = 1$.

## Appendix C. Isotropic compression of a spherical layer with combined viscoelastic and plastic flow

Isotropic compression of a hollow sphere is studied. The geometry is the same as in Appendices A and B. Consider the case where viscoelastic and plastic flows coexist in two different non-void domains of the spherical layer. Taking into account the tendency of decreasing shear stresses with radius $r$ at isotropic compression and the spherical symmetry, it is reasonable to assume that the boundary between the viscoelastic and plastic domains is a sphere of radius $R_{\rm c}$, $R_1 < R_c < R_2$, the inner spherical layer $R_1 < r < R_{\rm c}$ is the plastic domain, and the outer layer $R_{\rm c} < r < R_2$ is the viscoelastic domain. In the outer viscoelastic domain, general solutions obtained in Appendix B are applicable: (B3) for the maximum shear stress (B1) and (B5) for the normal radial stress. At the boundary of plastic domain, the maximum shear stress attains CRSS $\tau_{\rm c}$ resulting the following boundary condition:

$$\tau_{\max}\big|_{r=R_{\rm c}} = \tau_{\rm c} . \tag{C1}$$

The value of constant $A$ and the maximum shear stress distribution in the viscoelastic domain obtained from Eqs. (B3) and (C1) are

$$A = -\frac{2\Omega_n \tau_{\rm c}^n}{3\mu_0} R_{\rm c}^3 , \qquad \tau_{\max} = \tau_{\rm c} \left( \frac{R_{\rm c}}{r} \right)^3 . \tag{C2}$$

The value of constant $B$ and the distribution of $\sigma_{\rm rr}$ is obtained from the boundary condition for $\sigma_{\rm rr}$ at the outer boundary $r = R_2$ (A9) as

$$B = -\frac{4n}{3} \tau_{\rm c} \left( \frac{R_c}{R_2} \right)^{3/n} - p , \qquad \sigma_{\rm rr} = \frac{4n}{3} \tau_{\rm c} \left[ \left( \frac{R_{\rm c}}{r} \right)^{3/n} - \left( \frac{R_{\rm c}}{R_2} \right)^{3/n} \right] - p . \tag{C3}$$

In the plastic domain, momentum equation (B4) with plastic flow condition $\tau_{\max} = \tau_{\rm c}$ gives the differential equation for the radial normal stress,

$$\frac{{\rm d}\sigma_{\rm rr}}{{\rm d}r} + \frac{4\tau_{\rm c}}{r} = 0 . \tag{C4}$$

The solution of Eq. (C4) satisfying boundary condition for $\sigma_{\rm rr}$ at the inner boundary $r = R_1$ (A9) is

$$\sigma_{\rm rr} = -4\tau_{\rm c} \ln \frac{r}{R_1} . \tag{C5}$$

The value of $R_c$ is obtained from the condition of continuity of $\sigma_{rr}$ at $r = R_c$ resulting the following transcendent equation:

$$\ln\frac{R_c}{R_1} = \frac{p}{4\tau_c} - \frac{n}{3}\left[1 - \left(\frac{R_c}{R_2}\right)^{3/n}\right]. \tag{C6}$$

The deformation rate of the hollow sphere is

$$\frac{dR_2}{dt} = u_r\big|_{r=R_2} = \frac{A}{R_2^2} = -\frac{2\Omega_n\tau_c^n}{3\mu_0}\frac{R_c^3}{R_2^2}. \tag{C7}$$

Densification rate is

$$\frac{\dot{\rho}}{\rho} = -\frac{3}{R_2}\frac{dR_2}{dt} = \frac{2\Omega_n\tau_c^n}{\mu_0}\left(\frac{R_c}{R_2}\right)^3. \tag{C8}$$

Comparison of Eqs. (2.17) and (C8) gives the following expression for the bulk relaxation rate:

$$\omega_b = \Omega_n\frac{2K}{3\mu_0}\frac{\tau_c^n}{p}\left(\frac{R_c}{R_2}\right)^3, \tag{C9}$$

which is applicable if pressure $p$ is greater than

$$\frac{4n\tau_c}{3}(1-\theta^{1/n}), \tag{C10}$$

where $\theta$ is the porosity given by Eq. (B10). If the pressure becomes as low as (C10), the domain of plastic flow vanishes according to Eq. (C6), and Eq. (C9) reduces to Eq. (B9). Reasonably, Eq. (B9) becomes applicable if pressure $p$ is not greater than (C10) when no plastic flow occurs.

Equation (5.11) for the shear relaxation rate $\omega_{s0}$ is applicable in the viscoelastic domain. In the plastic domain, the explicit expression can be obtained from the constitutive relation (A2) as

$$\omega_{s0} = \Omega_n\tau_c^{n-1}\left(\frac{R_c}{r}\right)^3. \tag{C11}$$